\documentclass[%
reprint,
 amsmath,amssymb,
 aps,
]{revtex4-1}

\usepackage{graphicx}
\usepackage{dcolumn}
\usepackage{bm}
\usepackage{color}
\usepackage{amsmath}
\usepackage{comment}
\usepackage[utf8]{inputenc}
\usepackage{textcomp}
\usepackage{soul}
\usepackage{hyperref}
\usepackage{xr}

\usepackage{subfiles}
\usepackage{blindtext}
\usepackage{amsmath,amsfonts,amsthm,bm} 

\usepackage{float}
\newfloat{suppfig}{tbh}{losf}
\floatname{suppfig}{Supplementary Figure}

\begin{document}


\title{Spatial images from temporal data}

\author{Alex Turpin$^{1*}$, Gabriella Musarra$^1$, Valentin Kapitany$^1$, Francesco Tonolini$^2$, Ashley Lyons$^1$, Ilya Starshynov$^1$, Federica Villa$^3$, Enrico Conca$^3$, Francesco Fioranelli$^4$, Roderick Murray-Smith$^2$, Daniele Faccio$^1$}
\email{Corresponding author: \\ alex.turpin@glasgow.ac.uk; daniele.faccio@glasgow.ac.uk}

\affiliation{$^1$ School of Computing Science, University of Glasgow, Glasgow G12 8QQ, UK\\
$^2$ School of Physics \& Astronomy, University of Glasgow, Glasgow G12 8QQ, UK\\
$^3$Dipartimento di Elettronica, Informazione e Bioingegneria, Politecnico di Milano, 20133 Milano, Italy\\
$^4$Department of Microelectronics, TU Delft,  2628CD, Netherlands
}%

\date{\today}
             
\begin{abstract}

Traditional paradigms for imaging rely on the use of spatial structure either in the detector (pixels arrays) or in the illumination (patterned light). 
Removal of spatial structure in the detector or illumination, i.e. imaging with just a single-point sensor, would require solving a very strongly ill-posed inverse retrieval problem that to date has not been solved.
Here we demonstrate a data-driven approach in which full 3D information is obtained with just a single-point, single-photon avalanche diode that records the arrival time of photons reflected from a scene that is illuminated with short pulses of light. 
Imaging with single-point time-of-flight (temporal) data opens new routes in terms of speed, size, and functionality. As an example, we show how the training based on an optical time-of-flight camera enables a compact radio-frequency impulse RADAR transceiver to provide 3D images.
\end{abstract}

\maketitle

\section{Introduction}
The most common approach to image formation is obvious and intuitive: a light source illuminates the scene and the back-reflected light imaged with lenses onto a detector array (a camera). A second paradigm, single-pixel imaging, relies instead on the use of a single pixel for light detection, while the structure is placed in the illumination by spatially scanning the scene in some form 
\cite{2008:PRA:saphiro,duarte:2008:IEEE,padgett:2019:natphoton,wetzstein:2018:acm,radwell:2019:scirep,lyons:2019:CLEO}.  
Three-dimensional (3D) imaging can be obtained with these approaches gathering 
depth information 
via stereovision, holographic or time-of-flight techniques \cite{stereo,holography,padgett:2013:science,sun:2016:natcommun}.
In time-of-flight approaches the depth information is estimated by measuring the time needed by light to travel from the scene to the sensor \cite{book-lidar}. 
Many recent imaging approaches, ranging from sparse-photon imaging \cite{first_photon,morris,altmann} to non-line-of-sight imaging \cite{velten:2012:natcommun,gariepy:2016:natphoton,wetzstein:2018:nature,jin:2018:oe,jarabo:2017:oe,musarra:2019:arxiv}, also rely on computational techniques for enhancing imaging capabilities.
Among the various possible computational imaging algorithms \cite{faccio:2018:science}, machine learning \cite{jordan:2015:nature} and in particular deep learning \cite{lecun:2015:nature}, 
provides a statistical or data-driven model for enhanced image retrieval \cite{2019:ozcan:optica}. State-of-the-art deep learning techniques have been applied in computational imaging problems such as 3D microscopy \cite{waller:2015:nature,ozcan:2017:optica},
super-resolution imaging \cite{schechtman:2018:optica,ozcan:2019:natmeth}, phase recovery \cite{ozcan:2018:lsa,barbastathis:2018:prl}, lensless imaging \cite{barbastathis:2017:optica,ozcan:2018:acs} and imaging through complex media \cite{barbastathis:2018:optica,psaltis:2018:optica,tian:2018:optica,turpin:2018:oe,moser:2018:lsa,caramazza:2019:natcomms}.\\
%
All these imaging approaches use either a detector array or scanning/structured illumination to retrieve the transverse spatial information from the scene. This requirement
is clear if we consider the inverse problem that needs to be solved if data is collected only from a single point, non-scanning detector, and with no structure in the illumination: there are infinite possible scenes that could give the same measurement on the single point sensor thus rendering the inverse problem very strongly ill-posed. \\
Here, we introduce a new paradigm for spatial imaging based on single-point, 
time-resolving detectors. In our approach, the scene is flood-illuminated with a pulsed laser and the return light is focused and collected with a single-point single-photon avalanche diode (SPAD) detector, which records only the arrival time of the return photons from the whole scene in the form of a temporal histogram. During the measurement no spatial structure is imprinted at any stage either on the detector or the illumination source. Then, 
an artificial neural network (ANN) reconstructs the 3D scene from a single temporal histogram. We demonstrate 3D imaging of different objects, including humans, with a resolution sufficient to capture scene details and up to a depth of $4\,\rm{m}$. 
We prove that using the background of the scenes is a key element to detect, identify and image moving objects and we exploit it for our application.
Our approach is a conceptual change with respect to the common mechanisms for image formation, as spatial images are obtained from a single temporal histogram. This result lends itself to cross-modality imaging whereby training based on ground truth from an optical system can be applied to data from a completely different platform. As an example, we show that a single radio-frequency (RF) impulse RADAR transducer together with our ML algorithm is enough to retrieve 3D images.

\section{Single-point 3D imaging approach} 

Previous work has shown that, in addition to the object-sensor distance, the 3D profile of objects manifests through a particular temporal footprint that makes them classifiable even in cluttered environments \cite{caramazza:2018:scirep}. 
Here we extend this concept to full
imaging using only photon arrival-time from the scene. It is simple to construct a forward model 
where all points in the scene that are at some distance, $\mathbf{r}_i = (x_i,y_i,z_i)$, from the detector provide a related photon arrival time, $t_i = c^{-1} |\mathbf{r}_i| = c^{-1} \sqrt{x_i^2+y_i^2+z_i^2}$ (where $c$ is the speed of light).  
By recording the number of photons arriving at different times $t$, we can build up a temporal histogram that contains information about the scene in 3D. 

\begin{figure*}[!ht]
\centering
\includegraphics[width=0.8\linewidth]{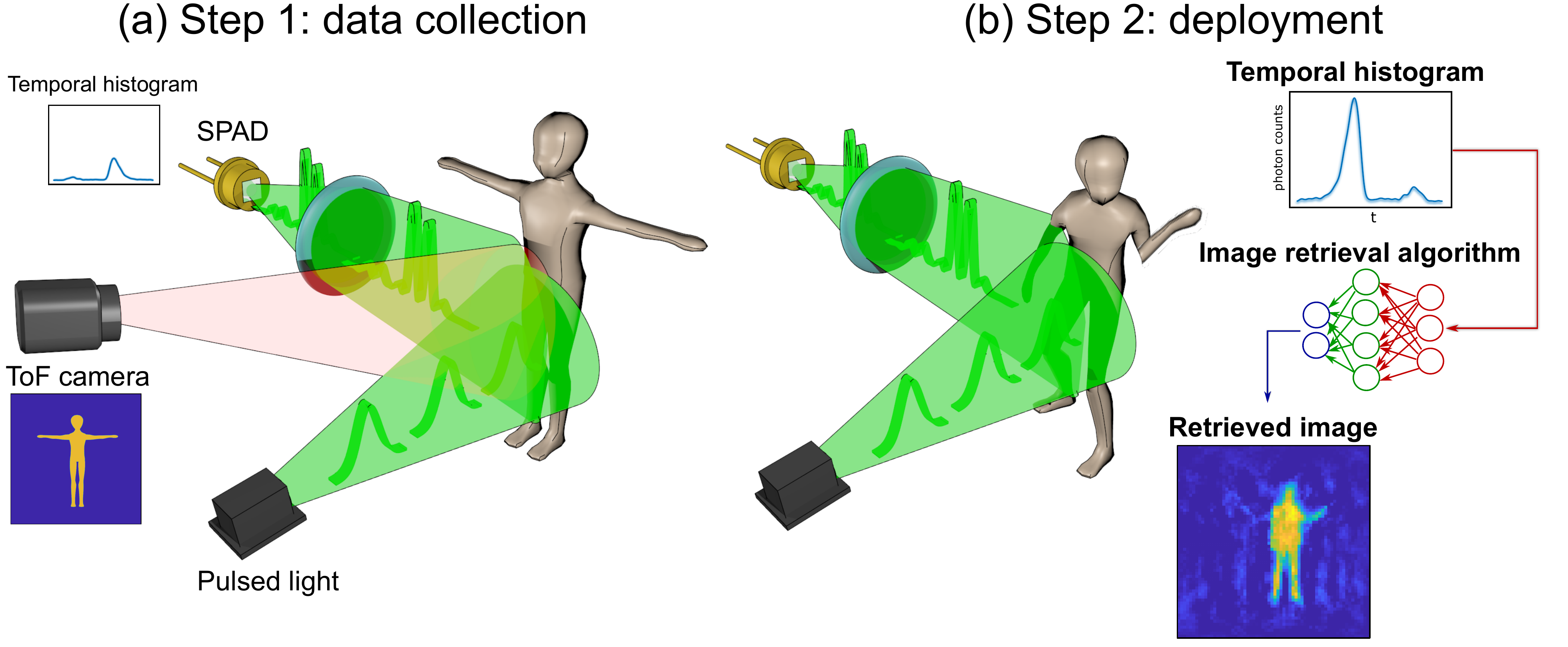}
\caption{
3D imaging with single-point time-resolving sensors. Our approach is divided into two steps: (a) a data collection step and (b) the deployment phase. During step 1 a pulsed laser beam flash-illuminates the scene and the reflected light is collected with a single-point sensor (in our case, a single-photon avalanche diode, SPAD) that provides a temporal histogram via time-correlated-single-photon counting (TCSPC). In parallel, a time-of-flight (ToF) camera records 3D images from the scene. The ToF camera operates independently from the SPAD and pulsed laser system. The SPAD temporal histograms and ToF 3D images are used to train the image retrieval ANN. Step 2 occurs only after the ANN is trained. During this deployment phase, only the pulsed laser source and SPAD are used: 3D images are retrieved from the temporal histograms alone.
}
\label{fig1}
\end{figure*}

However, solving the inverse problem is a much harder task. Indeed, obtaining the 3D coordinates $\mathbf{r}_i$ of objects the reflected photons from which contribute to a 1D temporal histogram (i.e. containing no spatial information in any form), is an extremely ill-posed problem. 
This data problem becomes even harder to solve when one realises that photons reflected from objects at coordinates placed within a spherical dome represented by the equation $(ct_i)^2 = x^2+y^2+z^2$ have the exact same arrival time $t_i$ at the detector and, as a consequence, they will contribute with equal probability to the same time bin on the histogram. 
Therefore, a single temporal histogram is not enough, in principle, to obtain a unique solution for the inverse problem and to retrieve meaningful shape or depth information that resembles the actual scene. This problem arises due to a lack of additional information, priors, or constraints on the scene (that is usually provided by using multiple light sources or detectors/pixels).
This lack of information and priors can be accounted for and brought into an image retrieval process in different ways. For instance, by following the methodology of common computational imaging algorithms, it is possible to have a forward model that generates different scenes compatible with the experimentally recorded temporal histogram and then use an iterative algorithm that estimates the degree of compatibility of these scenes with the data. However, if no prior information of the types of scenes is provided (e.g. imaging one or more humans continuously moving in an empty room) the number of solutions compatible with this approach is infinite and the algorithm would hardly converge towards the correct answer. 
In our work we take a different approach where this additional information is provided through priors based on data-sets containing the type of images that we aim to retrieve and a supervised machine learning algorithm that is trained for that purpose.

In more detail, the 3D imaging approach is depicted in Fig.~\ref{fig1} and consists of three main elements: i) a pulsed light source, ii) a single-point
time-resolving sensor, and iii) an image retrieval algorithm. The scene is flood-illuminated with the pulsed source and the resulting back-scattered photons are collected by the sensor. We use a single-point SPAD detector operated together with time-correlated single-photon counting (TCSPC) electronics
to form a temporal histogram [Fig.~\ref{fig1}(b)] from the photons arrival time
Objects placed at different positions within the scene and objects with different shapes provide different distributions of arrival times at the sensor \cite{caramazza:2018:scirep}. 

The histogram, $h$, measured by the single-point sensor can be mathematically described as $h = \mathcal{F}(S)$, where $S = S(\mathbf{r})$ represents the distribution of objects within the scene. The problem to solve is
the search for the function $\mathcal{F}^{-1}$ that maps the temporal histograms onto the scene. We adopt a supervised training approach (see Supplementary Material for details) 
by collecting, a series of temporal histograms corresponding to different scenes, together with the corresponding ground-truth 3D images collected with a commercial time-of-flight (ToF) camera. The ANN is then trained to find an approximate solution for 
$\mathcal{F}^{-1}$ and is finally used to reconstruct 3D images only from time-resolved measurements of new scenes that have not been seen during the training process. 
We recall that this is one of the key reasons for using a machine learning approach: once the algorithm has been trained (which happens only once), this can be used with unseen temporal histograms straight away, i.e. no further training is required. Moreover, the trained algorithm could be implemented on portable platforms for fast and lightweight applications, as it is extremely light computationally speaking.

\section{Numerical results}
\begin{figure}[!ht]
\centering
\includegraphics[width=0.7\linewidth]{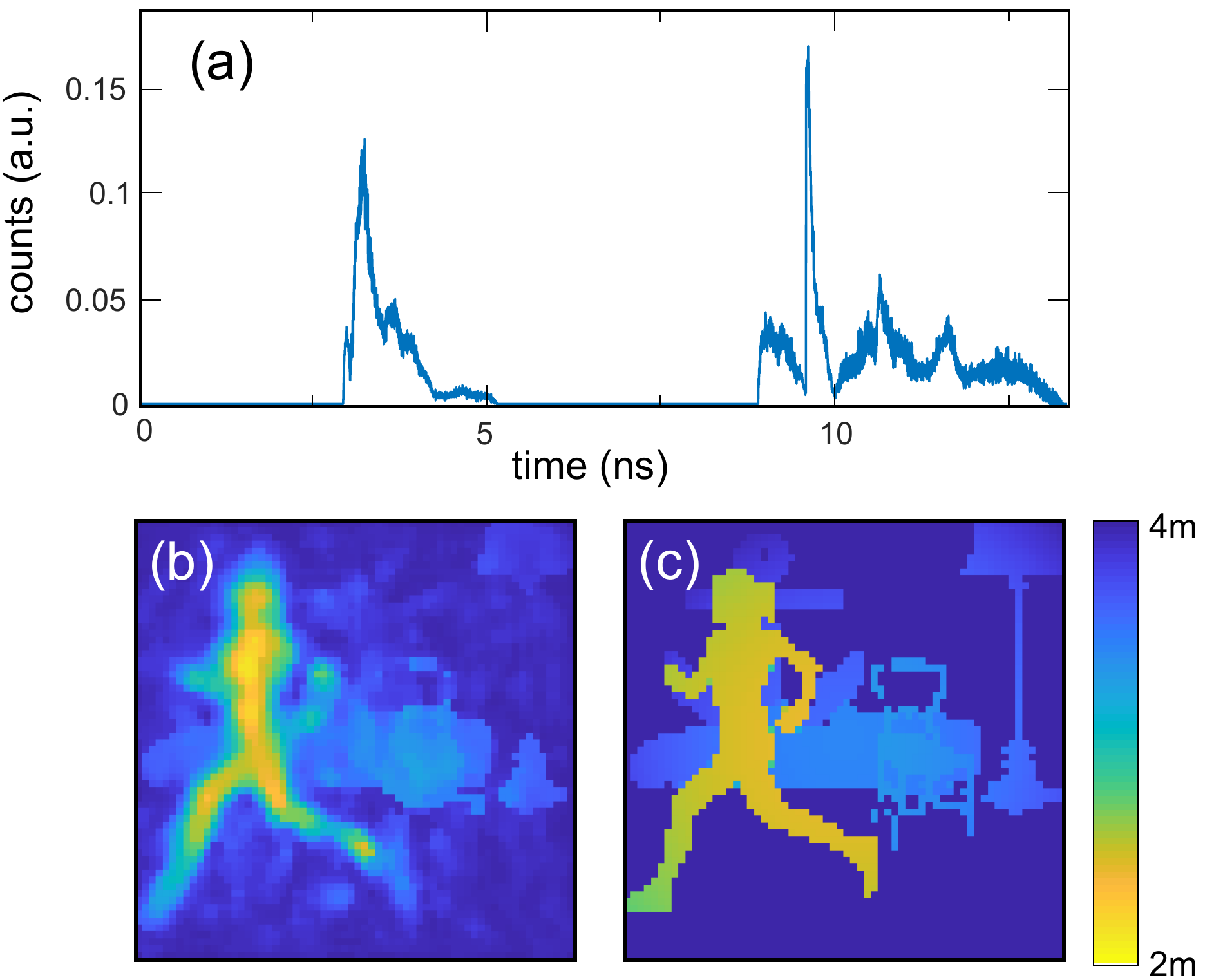}
\caption{
Numerical results showing 3D imaging from a single temporal histogram recorded with a single-point time-resolving detector. (a) Temporal trace obtained from the scene (shown in (c) as a color-encoded depth image). (b) 3D image obtained from our image retrieval algorithm when fed with the histogram from (a). The  colour bars describe the colour-encoded depth map. 
}
\label{fig2}
\end{figure}

To evaluate the validity of our approach, we first analyzed its performance with numerical simulations. We consider human-like objects with different poses, moving within a scene of $20\,\rm{m}^3$, which is represented as a color-encoded depth image, as shown in Fig.~\ref{fig2}(c).  
We assume 
flash illumination of the scene with a pulsed light source (with duration that is much shorter than all other timescales in the problem) and then calculate the photons arrival-time from every point of the scene. Simulating different scenes allows us to obtain multiple 3D images-temporal histograms pairs that are used to train 
the image retrieval algorithm (details about the structure of the ANN and training parameters can be found in Supplementary Material). 

Typical scenes consist of a static background with moving human figures in different poses, as shown in the Supplementary Material. After training the ANN, single temporal histograms are tested to reconstruct the related 3D scenes. To evaluate the potential performance in idealised conditions, for these simulations we assumed that the time bin width $\Delta t = 2.3\,\rm{ps}$ is also the actual temporal resolution (impulse response function, IRF) of the full system. 
The minimum resolvable transverse feature size or lateral object separation $\delta$ that can be distinguished with our technique depends on both the IRF, $\Delta t$, and the distance from the sensor, $d$:
\begin{equation}
    \delta (d, \Delta t) = c \Delta t \sqrt{\frac{2d}{c \Delta t} + 1},
    \label{eq1a}
\end{equation} 
where $c$ is the speed of light. In the depth direction, the spatial resolving power is determined only by the time-of-flight resolution (as in standard LiDAR), i.e. $\delta_z=c\Delta t$. At a distance of $4\,\rm{m}$ from the detector, for $\Delta t=2.3\,\rm{ps}$ we can expect a transverse image resolution of $7\,\rm{cm}$, which will degrade to $77\,\rm{cm}$ for $\Delta t=250\,\rm{ps}$. The impact of the latter realistic time-response will be shown in the following experimental results. Figure~\ref{fig2}(a) shows one example of a temporal histogram constructed from the scene in Fig.~\ref{fig2}(c). Figure~\ref{fig2}(b) shows the scene reconstructed using the numerically trained ANN and highlights the relatively precise rendition of both depth and transverse details in the scene. 

\section{Experimental results}
\subsection{Optical pulses}
After numerically demonstrating the concept of 3D imaging with single-point time-resolving detectors, we test its applicability in an experimental environment. We flood-illuminate the scene with a pulsed laser source at $(550 \pm 25)\,\rm{nm}$, with pulse width of $\tau = 75\,\rm{ps}$. Our scenes are formed by a variety of fixed background objects, up to a depth of $4\,\rm{m}$ (the maximum distance allowed by our ToF camera), while different additional objects  (people, large-scale objects) are moving dynamically around the scene. 
Although in our experiments we use a ToF camera, we note that any other 3D imaging system, such as LiDAR, stereoimaging, or holography devices could be used for collecting the ground truth data for the training process. The ToF camera is synchronized with a SPAD detector equipped with TCSPC electronics that provide temporal histograms from the back-reflected light that is collected from the whole scene with an angular aperture of $\sim30^{\circ}$ and an IRF $\Delta t=250\,\rm{ps}$  (measured with a small $2\,\rm{cm}$ mirror in the scene). 

\begin{figure*}[!ht]
\centering
\includegraphics[width=0.85\linewidth]{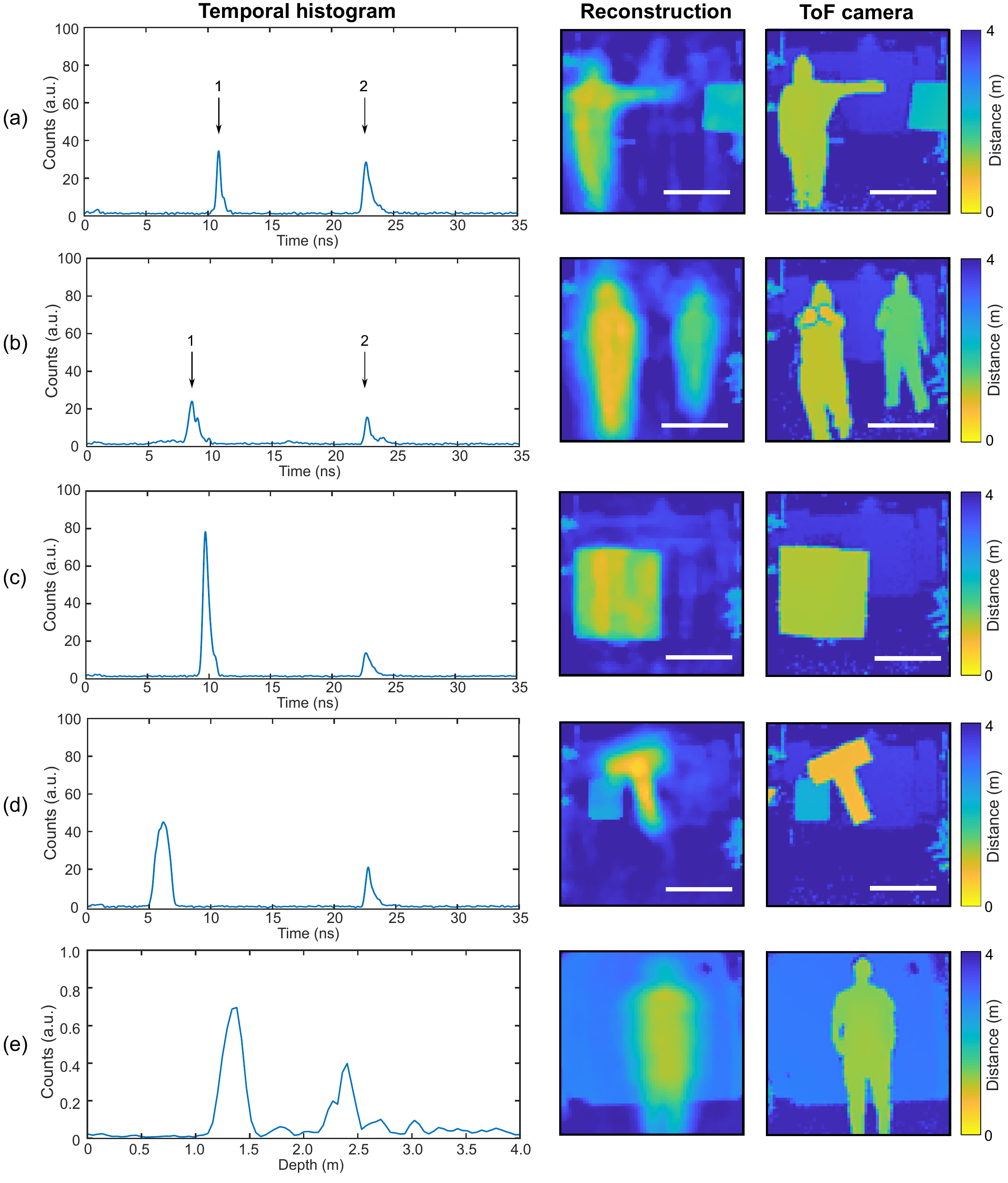}
\caption{
Experimental results showing the performance of our system recovering the 3D image from temporal histograms in different scenarios. The first column shows temporal histograms recorded with the SPAD sensor and TCSPC electronics [rows (a)-(d)] or with the RADAR transceiver [row (e)], while the last column represents 3D images measured directly with the ToF camera for comparison to the reconstructed images (second column). The colour bars describe the colour-encoded depth map. The white scale bar corresponds to $80\,\rm{cm}$ at $2\,\rm{m}$ distance. Full videos are available in the supplementary information (Supplementary Videos 1 and 2).
}
\label{fig3}
\end{figure*}

In Fig.~\ref{fig3} the first column shows examples of recorded temporal histograms, the second column shows the images reconstructed from these temporal histograms, and the last column shows the ground truth images obtained with the ToF camera, for comparison. 
Full movies with continuous movement of the people and objects within the scenes, acquired at $10\,\rm{fps}$, are shown in Supplementary Video 1.
As can be seen, even with the relatively long IRF of our system, it is possible to retrieve the full 3D distribution of the moving people and objects within the scene from the single-point, temporal histograms. 
Compared to the numerical results, the larger IRF leads to the loss of some details in the shapes, such as arms or legs that are not fully recovered.

We can see these limitations for example  in the reconstruction of the letter `T', row (d) of Fig.~\ref{fig3} (with dimensions $39 \times 51\,\rm{cm}^2$) and especially in the Supplementary Video 1, where the algorithm is able to detect the object, but struggles to obtain the correct shape. This lateral resolution power [Eq.~(\ref{eq1a})] would be improved for example to  $25\,\rm{cm}$ with $\Delta t=25\,\rm{ps}$ and increases with a square-root law with distance, implying a relatively slow deterioration versus distance (for example, resolution would be $50\,\rm{cm}$ at $20\,\rm{m}$ distance).

Specific shape information is retrieved from all features in the scene, both dynamic (e.g. moving people) and static (e.g. objects in the background). This can be seen in Fig.~\ref{fig3}(a) and Fig.~\ref{fig3}(b), where both temporal histograms have peaks that are placed at similar positions, yet the reconstructed scenes are different. On the one hand, in Fig.~\ref{fig3}(a) the ANN recognizes the box at the right of the image (corresponding to the peak 2 in the histogram) as a static background object that was present in all of the training data, while the person (peak 1) is identified as a dynamic object with a certain shape given by the peak structure. In contrast, the ANN recognizes both temporal peaks 1 and 2 in  Fig.~\ref{fig3}(b) as people moving dynamically through the scene (as these were not constant/static in the training data). 

This example highlights the role of the background, which 
is key also in removing ambiguities that would arise in the presence of a single isolated object moving in front of a uniform background (for instance, a single isolated histogram peak could be interpreted as a person placed either to the left or to the right of the scene - see Supplementary Material for details). 

\subsection{GHz pulses}
We further analyze the impact of the IRF on the image quality reconstruction by repeating the reconstruction with temporal histograms that are convolved with Gaussian point-spread functions with different temporal widths. The results (Supplementary Material), show that although longer IRFs  degrade image reconstruction, the shape and 3D location of people in the scene are still easily recognisable even when using an IRF of $1\,\rm{ns}$ (corresponding nominally to a lateral spatial resolution of $\approx 1.4\,\rm{m}$ at $4\,\rm{m}$ distance). This opens the possibility for cross-modality imaging, understood here in the context of detecting signals in one modality or domain and extracting information from a completely different modality. In particular, our approach offers new perspectives for sensing with pulsed sources outside the optical domain, which typically have ns-scale IRF, and providing 3D images with resolutions typical of those obtained in the optical domain e.g. with the Tof camera. To demonstrate this, we replaced the pulsed laser source, SPAD detector, and other optical elements with an impulse RADAR transceiver emitting  at $7.29\,\rm{GHz}$ (see Supplementary Material and Supplementary Video 2). After re-training the ANN using the RADAR transceiver for the time-series data and the optical ToF camera for the ground truth images, we can retrieve a person's shape and location in 3D from a single RADAR histogram [see Fig.~\ref{fig3}(e)], thus transforming a ubiquitous RADAR distance sensor into a full imaging device. 

\section{Discussion}

Although the maximum resolution of the reconstructed images is limited by the resolution of the 3D sensor used during training, overall, the quality of the final image is essentially determined by the temporal resolution of the single-point time-resolving detector. With state-of-the-art sensors currently heading towards $10\,\rm{ps}$ or better resolution, there is potential for 3D imaging with spatial resolution better than $10\,\rm{cm}$ at distances of $10\,\rm{m}$ or more. The precision in the image reconstruction is also determined by the reconstruction algorithm with improvement possible by using more advanced algorithms (including non-ML-based ones) and also fusing the ToF flight data with other sensor data, e.g. a standard CCD/CMOS cameras. 
However, using single-point SPADs has promising potential for high-speed implementations. After the algorithm is trained (which is performed only once), the image reconstruction problem has two different time-frames: (i) the algorithm reconstruction time and (ii) the histogram data collection time. On the one hand (i) is easy to be measured with the computer directly, providing times on the order of $10-30\mu\rm{s}$ for the algorithm here used. On the other hand, to account for (ii) different factors need to be considered. First, typical TDCs run at 10~MHz. Our experiments indicate that about 1000 photons per temporal histogram are needed to retrieve a meaningful image, which leads to histogram recording frame-rates of $\approx$ 10~kHz, which is reduced to 1~kHz or less if we are in the photon starved regime and we account for data transfer to an electronics board. This frame rate could be increased further if instead of a single pixel, a SPAD array is used as a ``super-pixel'' by adding all the outputs into a single histogram, thus collecting hundreds of photons for each illumination pulse (e.g. with a 32$\times$32 array).
Such devices are commercially available and can run at 100~kHz, which would therefore define the rate at which we could collect single temporal histograms  These estimates indicate a clear potential for imaging at 1-100 kHz with no scanning parts and a retrieval process that can match this rate even when running with standard software and hardware.

Although the above-discussed advantages of our approach for imaging in terms of data processing and hardware are important, the key message in this work is the potential of using temporal data gathered with just a single pixel for spatial imaging. This approach broadens the remit of what is traditionally considered to constitute image information. The same concept is therefore transferable to any device that is capable of probing a scene with short pulses and precisely measuring the return ``echo'', for example RADAR and acoustic distance sensors, indicating a different route for achieving for example full $360^{\circ}$ situational awareness in autonomous vehicles and smart devices or wearable devices.

\section{Conclusions}
Current state-of-art imaging techniques use either a detector array or scanning/structured illumination to retrieve the transverse spatial information from the scene, i.e. they relate spatial information directly to some type of spatial sensing of the scene. In this work we have demonstrated an alternative approach to imaging that suppresses all forms of spatial sensing and relies only on a data-driven image retrieval algorithm processing a single time-series of data collected from the imaged scene.
The experiments were carried out in scenes where objects were moving in front of a static background. This makes our approach well suited for applications where the device needs to be placed at a fixed position during operation, i.e. with a fixed background. 
There are multiple situations where operating in a fixed environment is useful. Examples are surveillance and security in public spaces, etc. These are examples where the background (e.g. walls of the room, buildings) do not change at all and they are also very widespread scenarios. Currently, cities have spaces that are constantly monitored with CCTV cameras that also potentially record information from which it is possible to extract information that breaches data protection policies. Our approach is therefore highly indicated for cases where one requires human activity in a fixed area and in a data-compliant way. The approach shown here would be also valid in a slowly changing environment, where training could in principle be continuously updated. Indeed, background objects will appear static if they change at a slower rate (and/or are at a larger distance) with respect to the dynamic elements of the scene or slower than the acquisition rate of the sensor. An interesting route for future research is of course to also investigate methods that account for dynamic backgrounds.

Finally, an interesting extension would be to non-line-of-sight (NLOS) imaging, especially given the latest developments exploiting computational techniques for image information retrieval from temporal data \cite{NLOSreview,gariepy2016detection,o2018confocal,liu2019non,IseringhausenNLOS2020} and the availability of public data-sets \cite{galindo19-NLOSDataset,hullin-dataset}.

\section*{Acknowledgements}
DF is supported by the Royal Academy of Engineering under the Chairs in Emerging Technologies scheme. AT acknowledges funding to the Alexander von Humboldt Stiftung. DF and RMS acknowledge financial support from EPSRC (UK, grant no. EP/T00097X/1). FT acknowledges funding from Amazon.

\section{Simulating time-of-flight data}
\begin{figure*}
\centering
\includegraphics[width=0.75\linewidth]{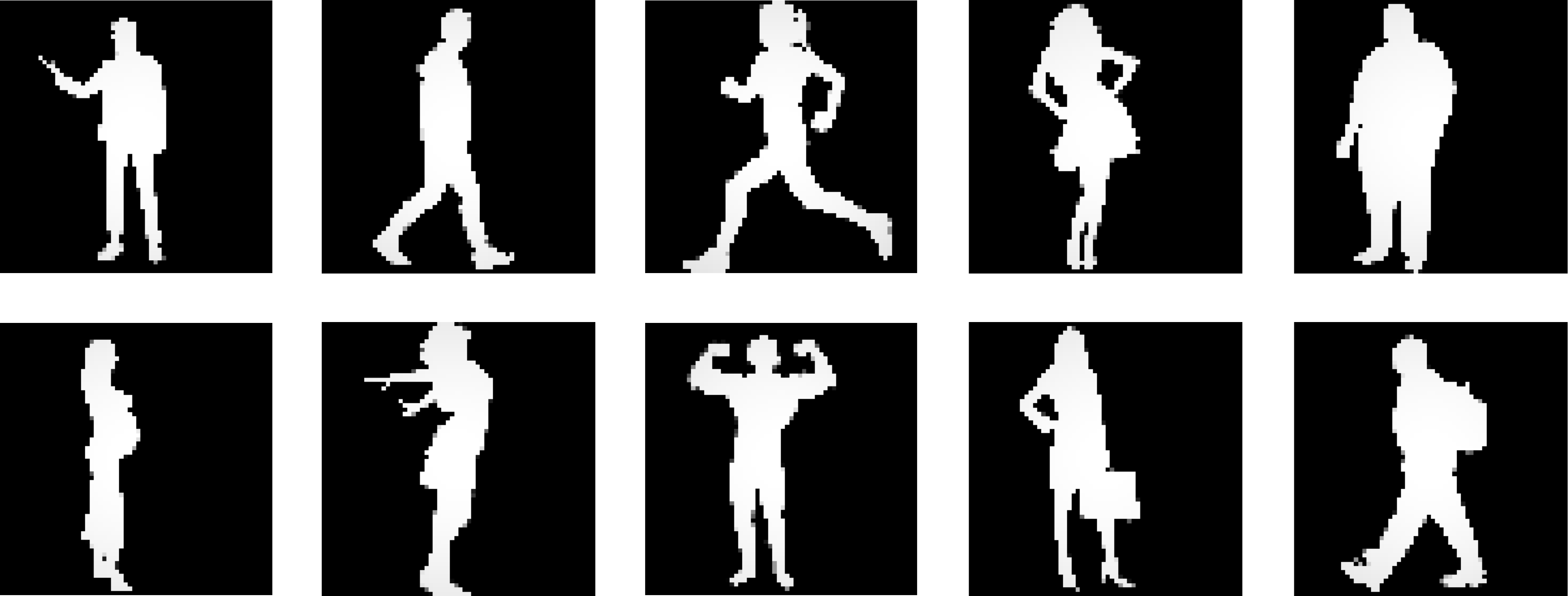}
\caption{
Human silhouettes  with different shapes and poses used to generate the data-set used to train the algorithm. 
}
\label{sf1}
\end{figure*}

To test our approach under the best ideal conditions, we performed numerical simulations where we generate synthetic scenes containing flat human-like silhouettes in different poses, as shown in Fig.~\ref{sf1}. We perform data augmentation of this data-set, originally consisting of 10 different humans and poses, by mirroring each image in the horizontal direction, and by also translating the humans around the scene in all directions $(x,y,z)$. This allows to augment the data set from the 10 original images to 4000 images used to train our neural network. In order to simulate the expected scaling factor of objects when increasing its distance from the observation point, we apply a size scaling factor to the image, depending on its coordinates $(x_i,y_i,z_i)$: 
\begin{equation}
    S = \frac{2}{d},
\end{equation}
where $d = \sqrt{(x_i^2+y_i^2+z_i^2)}$. The field of view considered in our simulations forms  we have assumed a viewing angle of $52^{\circ}$ from the virtual 3D camera. 
Each human silhouettes is translated to 200 different positions across the field of view of the scene: 10 different depths $z$ and 20 in $x$. This, together with the horizontal mirroring of the individual, gives us a total data set of 400 scenes per figure (4000 in total). 

\begin{figure*}
\centering
\includegraphics[width=0.9\linewidth]{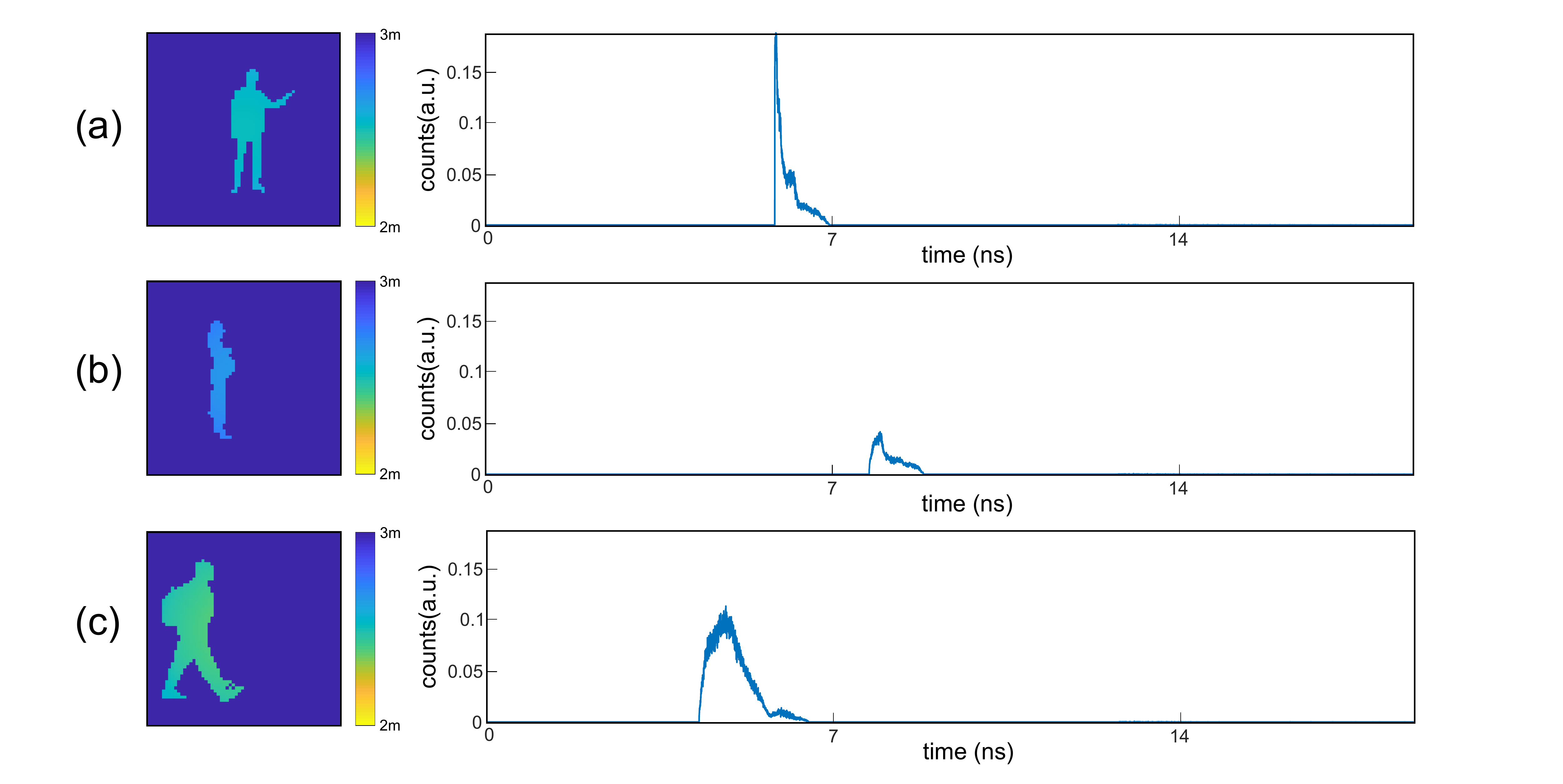}
\caption{
Examples of 3D images (depth is encoded in color) (left column) and corresponding temporal histograms (right column) from some of the data used to train the algorithm. 
}
\label{sf2}
\end{figure*}

We then assume that we illuminate the scene with a pulsed laser and estimate the arrival time of photons coming from each pixel of each image of the data set. For a given voxel in the image corresponding to a spatial position $\mathbf{r_i} = (x_i,y_i,z_i)$, its time of flight is $t_i = (2c)^{-1} \sqrt{x_i^2+y_i^2+z_i^2}$. From this information, we transform our data set into 3D images, where the value of every pixel encodes the time of flight, i.e. we generate a color-encoded depth 3D image, as shown in Fig.~\ref{sf2}. As we assume uniform reflectivity for all objects, the number of photons $n_i$ back-reflected from every point on the scene is inversely proportional to the voxel distance to the origin, i.e. $n_i \propto P_0 / \| \mathbf{r_i} \|^4$, where $P_0$ is the laser power. This information allows us to create a temporal histogram from the scene indicating how many photons arrive at a certain time to the virtual detector, see Fig.~\ref{sf2}. We thus generate our 3D image-temporal histogram pairs that are required for training of the artificial neural network (ANN). 

\begin{figure*}
\centering
\includegraphics[width=0.85\linewidth]{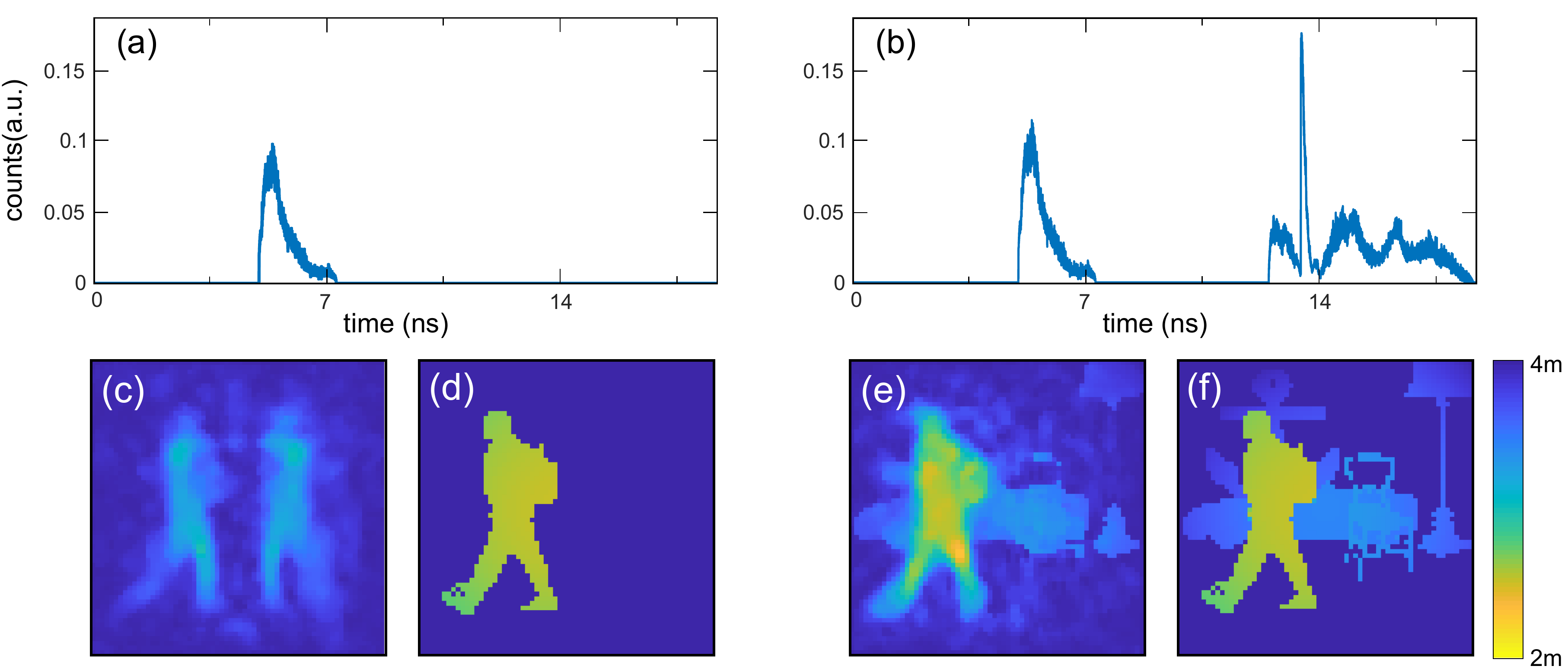}
\caption{
Reconstruction performance of different scenarios with objects permanently placed at certain positions within the scene. In the absence of background [(a), (c), (d)], the temporal trace of the individual alone is not enough to fully reconstruct its shape and position correctly. In a more natural scenario [(b), (e), (f)], with objects placed in the background, the temporal trace contains enough information for successful 3D imaging.
}
\label{sf3}
\end{figure*}

In a general scenario, it is very unlikely to find such clean backgrounds as the ones observed in Fig.~\ref{sf2}, where only the human individuals appear in scene. Therefore, to make our simulations more realistic we also added some objects to the background to mimic, for instance, what one would expect when walking into a room. We performed simulations with two different background scenes: a totally uniform (empty) background [Figs.~\ref{sf3}(a), (c), and (d)], and a background containing some common objects [Figs.~\ref{sf3}(b), (e), and (f)]. Figures~\ref{sf3}(a) and (b) are the temporal histograms corresponding to their respective ground-truth scenes, i.e. Figs.~\ref{sf3}(d) and (f). The predicted 3D scenes from our image retrieval algorithm are shown in Figures~\ref{sf3}(c) and (d). Note that in the absence of background objects (left-hand-side block), the algorithm struggles to reconstruct the 3D scene properly as the number and different shape of objects compatible with a particular temporal histogram is high. In particular, there is an evident left-right symmetry due to the fact that any object and its mirror image are both compatible with the same temporal trace. This symmetry is removed when including background objects (right-hand-side block), as the human silhouette will both add varying signals to the temporal for different positions, while subtracting signal from the corresponding background peaks observed when the silhouette is not present. Given that the subtracted signal data depends on the exact shape and location of the silhouette, the background plays a crucial role in solving the otherwise ill-posed problem and provides indirect information of the silhouette's absolute position that is learned by the ANN during the training process.\\
For the simulations shown in the main document, we used the background corresponding to Figs.~\ref{sf3}(b), (e), and (f).

\section{Image retrieval algorithm}
\begin{figure*}
\centering
\includegraphics[width=0.5\linewidth]{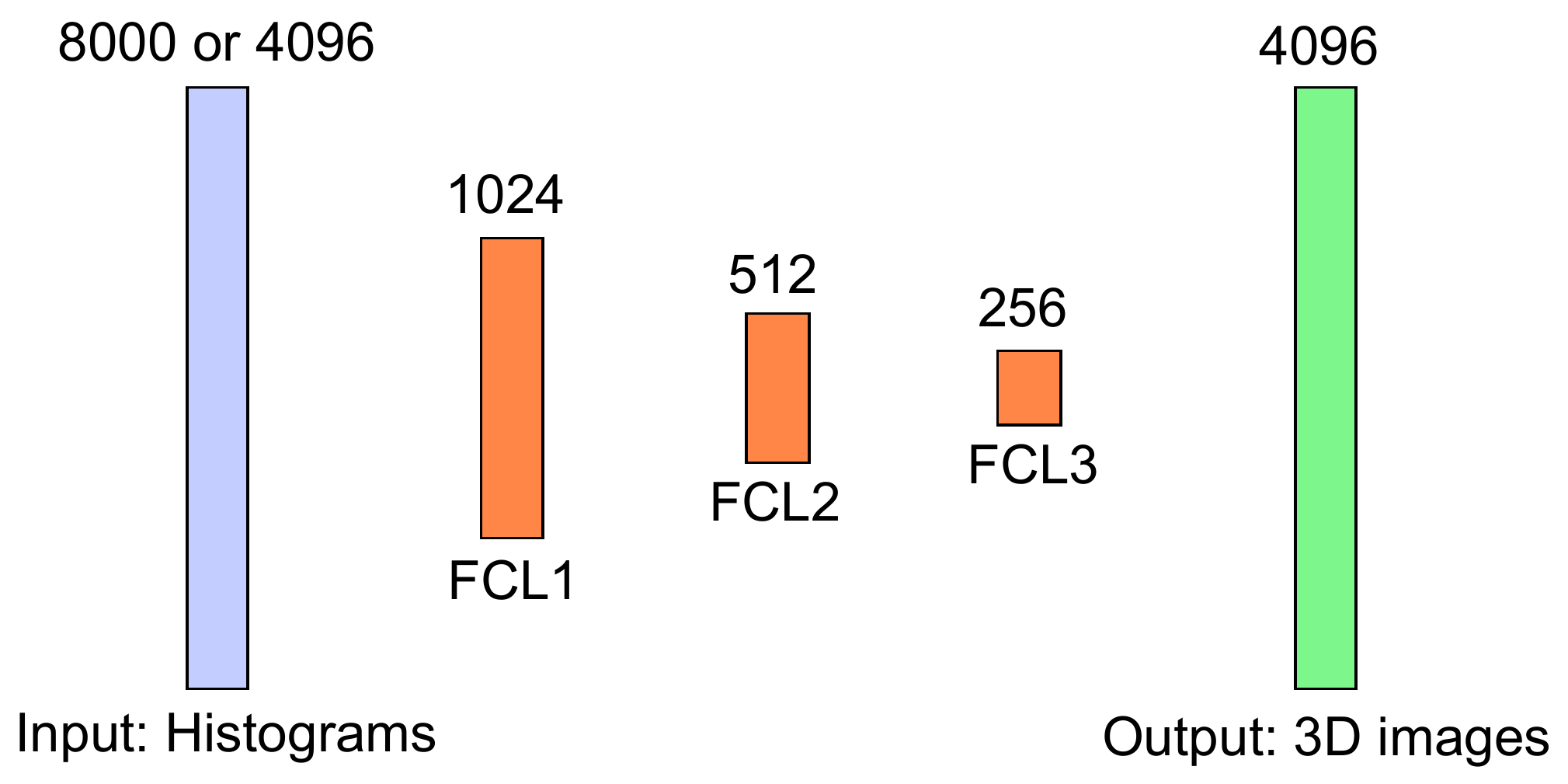}
\caption{
Sketch of the artificial neural network used: a multi-layer perceptron consisting of one input layer (with 8000 nodes), one output layer (with 4096 nodes), and three hidden fully-connected layers (FCL, with 1024, 512, and 256 nodes respectively). After each layer hidden layer we apply a $\tanh$ activation function (not shown in the figure). The loss used is the mean square error. }
\label{sf4}
\end{figure*}

Given a histogram $h = \mathcal{F}(S)$ recorded by the single-point time-resolving detector from the scene $S = S(\mathbf{r})$, the task of our algorithm is to find the function $\mathcal{F}^{-1}$ that maps $h$ onto $S$. A way to overcome this highly unconstrained inverse problem is by using prior information on the objects in the 3D scenes. To this end, we make use of a supervised training approach where we train an artificial neural network with pairs of temporal histograms and 3D images. Both the temporal histogram and the 3D images are treated as vectors with corresponding dimensions $1 \times 8000$ and $64 \times 64 = 1 \times 4096$ for the simulated data and $1 \times 1800$ and $64 \times 64 = 1 \times 4096$ for the experimental data (the details of the experiment are given in the next section). 
We implemented a multilayer perceptron (MLP) \cite{goodfellow:2016:book}. Perceptrons are mathematical models that produce a single output from a linear combination of multiple weighted real-valued inputs after passing through a nonlinear activation function. Mathematically, this operation can be expressed as: 
\begin{equation}
    \mathbf{y} = \phi \left( \mathbf{w}^T \mathbf{x} + b \right),
    \label{perceptron}
\end{equation}
where $\mathbf{y}$ is the output vector, $\mathbf{x}$ is the input vector, $\mathbf{w}$ is the weight matrix, and $b$ is the bias used to shift the threshold of the activation function $\phi$ (a $\tanh$ in our case). Our MLP, summarized in Fig.~\ref{sf4}, is composed of five layers: an input layer (temporal traces, having 8000 or 1800 nodes), 3 hidden fully-connected layer (with corresponding number of nodes: 1024, 512, 256), and an output layer (3D images, 4096 nodes).

Our choice of the algorithm type and structure is motivated by the physics of our problem: every time bin of the time histogram has contributions from all regions of the scene within a certain time of flight and, as a consequence, it contributes to multiple regions of the image during reconstruction. This is exactly what perceptrons do: connecting every point of the input with every point of the output data, see \eqref{perceptron}. Adding multiple perceptron layers in cascade increases the complexity and the flexibility of the algorithm, while modifying the dimensions of these layers allows the algorithm to concentrate on certain features of the histogram. We therefore experimented with different number of layers and different dimensions until we found a good compromise between training times, performance, and simplicity.

By training the MLP on sets of temporal histograms - 3D images pairs we learn an approximate function for the mapping $\mathcal{F}^{-1}$ which allow us to reconstruct 3D images from a single temporal histogram. Examples of the type of data used to train the algorithm are given in Figs.~2 and 4 in the main document. For the numerical/experimental results, 1800/9000 and 200/1000 pairs are used for training and testing, respectively, whilst a 7\% of the training data was used for validating. During experiments, the total 10000 ToF-image - temporal-histogram pairs were obtained within a single data acquisition sequence. However, we note that the data used for testing was never used during training and that this data also belongs to different periods of the data acquisition process during experiments. We also note that the only processing carried out on our data was a normalisation to the range $[0, 1]$ before passing through the algorithm. There is certainly room for future development in terms of data processing, e.g. time-gating or time-correlation between sequences, to improve the imaging capabilities of the algorithm. During learning, the cost function minimised for each image pixel, $i$, is the mean square error, $MSE$ 
\cite{keras}: %
\begin{equation}
\textit{MSE}_i = (y_i - s_i)^2,
\end{equation}
where $s_i$ and $y_i$ are the measured and predicted values of the depth of the pixel $i$ in the scene $S$, respectively. 
With the limitations in terms of power provided by our laser, we did not find any restrictions to train our algorithm within the 1-4m depth range. 
The MLP is implemented in TensorFlow \cite{tensorflow2015} using Keras \cite{keras} and training is performed by the Adam optimizer on a desktop computer equipped with a Intel Core i7 Eight Core Processor i7-7820X at $3.6\,\rm{GHz}$ and a NVIDIA GeForce RTX 2080 Ti with $11\,\rm{Gb}$ of memory. The training time depends on the number of images used, the batch size (number of images taken for each iteration of the training algorithm, 64 in our case), and the number of epochs (200), and requires  $26\,\rm{min}$ in total. After training the algorithm, this can recover a 3D image from a single temporal histogram in $30\,\mu\rm{s}$ on a standard laptop. 

\section{Experiment details}

In a first step we collect pairs of co-registered temporal histograms and ToF camera measurements. We used a supercontinuum laser source (NKT SuperK EXTREME) delivering pulses with $75\,\rm{ps}$ pulse duration and average power of $250\,\rm{mW}$ after a $50\,\rm{nm}$ band-pass filter centered at $550\,\rm{nm}$. A 10x microscope objective (Olympus plan achromat) with $\rm{NA} = 0.25$ and $\rm{WD} = 10.6\,\rm{mm}$ is used to flood-illuminate the scene with an opening angle of $\approx 30^{\circ}$. This opening angle ensures an illumination circle with diameter of $2.15\,\rm{m}$ at $4\,\rm{m}$ distance. The laser beam is expanded providing a 4x magnification that fills the back aperture of the objective and the IR radiation from the laser pump was filtered with two IR mirrors. Light scattered by objects placed inside the illumination cone is collected by a second microscope objective (40 $\times$, Nikon plan fluor, $\rm{NA} = 0.75$, $\rm{WD} = 0.66\,\rm{mm}$) and focused on a $50\times50\,\mu\rm{m}^{2}$ area SPAD sensor that retrieves the temporal trace of the scattered light by means of TCSPC electronics \cite{sanzaro:2018:IEEE}.
During the training process, we use a ToF camera (flexx PMD Technologies) with a depth range of $0.1-4\,\rm{m}$ and a maximum resolution of $224 \times 171$ pixels that is triggered via software together with the time-recording electronics. The ToF images are cropped and down-sampled to a resolution of $64 \times 64$ for training the ANN. After the ML algorithm is trained, it is tested with new data taken from new scenes and hence new temporal histograms  from the scene. 

\begin{figure*}
\centering
\includegraphics[width=0.5\linewidth]{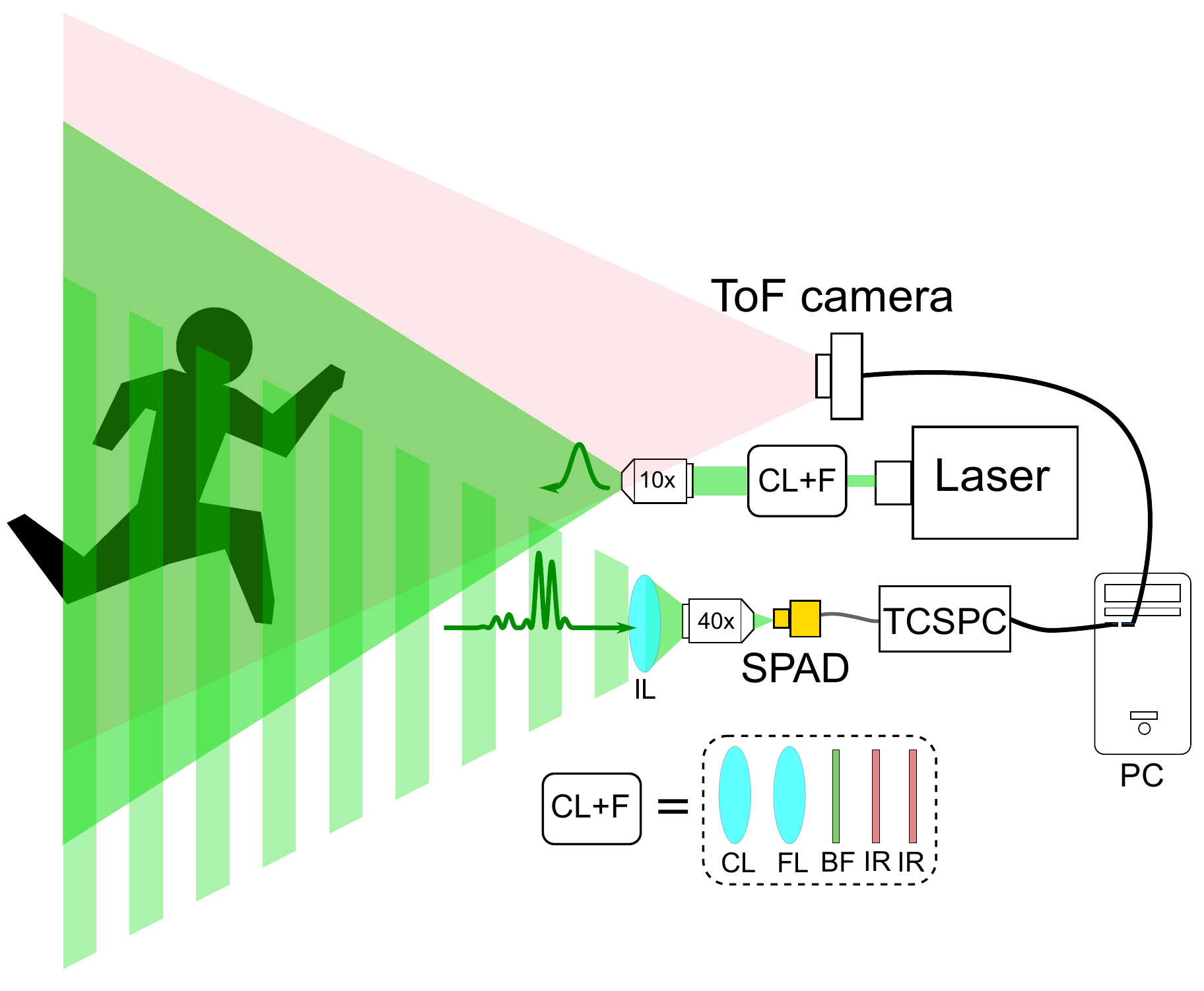}
\caption{
Detailed experimental set-up, see text for more details. A laser beam at $(550 \pm 50)\,\rm{nm}$ delivering pulses with $\tau = 75\,\rm{ps}$, at a power of $250\,\rm{mW}$ is expanded with a 10x microscope objective to flood-illuminate the scene, providing an opening angle of $30^{\circ}$. A pair of lenses used as 4x beam expander are used to fill the back aperture of the microscope objective, while two IR mirrors are used to filter out the laser pump. The back-scattered light from the scene is collected with a lens with focal $60\,\rm{mm}$ and a 40x objective that concentrates light onto the single-pixel SPAD, that retrieves temporal histograms via TCSPC. In parallel, a ToF camera grabs 3D images of the scene that are used as ground truth during training. }
\label{sf5}
\end{figure*}

In our proof-of-principle experiments we trained the system with two different individuals (either separately or  both present at the same time) moving around the scene, and also with non-human shapes (such as the letter `T' and the square shown in the main document) that we moved through the scene, also changing their orientations. To provide the algorithm with more variability, we also changed the position of some items in the background of the scene during the data acquisition process. In order to fix the reflectivity of the moving objects, we used a white overall suit for humans and white cardboard for the shapes.

The temporal impulse response function (IRF) of our system depends on the time resolving electronics and the pulse length. We directly measured the IRF by placing a mirror with diameter of $25.4\,\rm{mm}$ reflecting light back to the SPAD sensor. The IRF of the system is taken as the full width at half maximum of the time histogram, which was measured to be $250\,\rm{ps}$. 
In all our experiments we interfaced the SPAD sensor and TCSPC electronics, RADAR chip, and ToF camera via MATLAB. This allowed us to record temporal histograms at a rate of $10\,\rm{Hz}$ and $1\,\rm{Hz}$ for the SPAD and RADAR chip, respectively. 

For the cross-modality RADAR imaging, we used an ultra wide band impulse RADAR chip (Novelda XeThru X4). The chip has a single transmitter and receiver channel,  operates at $(7.29 \pm 1.4)\,\rm{GHz}$, has a pulse duration of $670\,\rm{ps}$, and a sampling rate of $23 \times 10^9$ samples/s.  

\section{Quality of the reconstructed images}

To quantify the quality of the reconstructed images we have compared these with their corresponding 3D images recorded with the ToF camera through the structural similarity index (SSIM). The SSIM is a human perception-based measure that compares the image content by separating the contributions of luminance, contrast, and structure (see \cite{simoncelli:2004:IEEE} for details). The SSIM takes values within the range $[-1,1]$, where 1 corresponds to two identical sets of data and -1 means that both images have no similar structure.

\subsection{SSIM of experimental data}

\begin{figure*}[htb]
\centering
\includegraphics[width=0.8\linewidth]{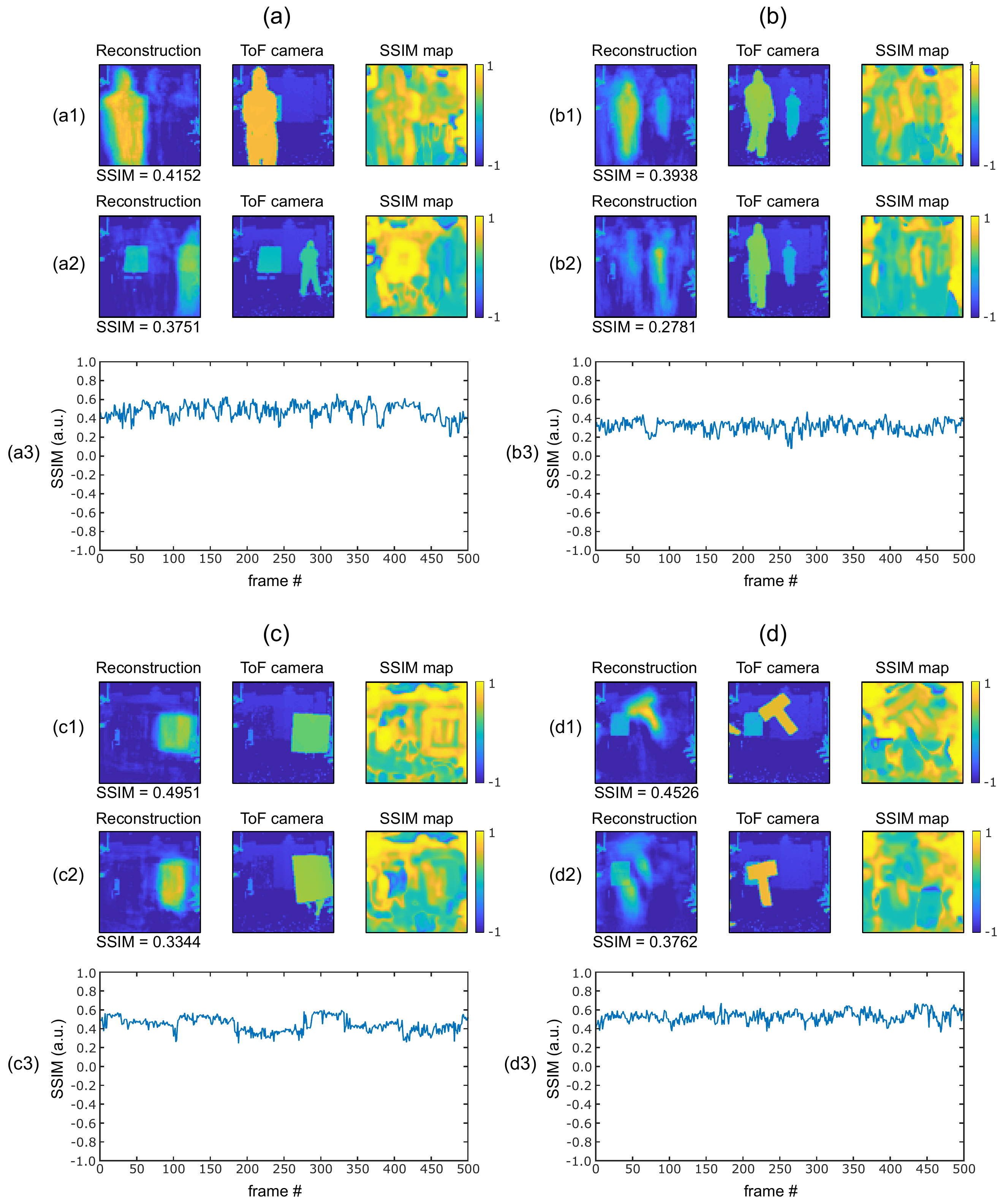}
\caption{
Quality of the reconstructed images computed through the Structural Similarity Index (SSIM). Each block (a) – (d), depicts respectively the cases with one person, two people, a square, and the letter T; and it is divided in three sections n$i$ (with n = a,b,c,d, $i = 1,2,3$). Sections n$i$ with $i = 1,2$ show 3D images obtained with our algorithm at the left ($i = 1$ for good reconstructions and $i = 2$ for worse reconstructions), the ground truth at the centre, and the SSIM maps between ground truth and reconstruction at the right. n3 (with n = a,b,c,d) shows a plot of the mean value of the SSIM maps for 500 different images  (indicated as a "frame $\#$" as each image corresponds to a frame in a recorded video sequence).
}
\label{sf9}
\end{figure*}

We first quantify the quality of the reconstructed images from temporal histograms recorded experimentally. Fig.~\ref{sf9} shows the results and the SSIMs for the experiments using pulsed light for the cases of (a) one person, (b) two people, (c) a square, and (d) the letter T. Each box (n) (with n = a,b,c,d) is divided in three sections n$i$ (with $i = 1,2,3$). Sections n$i$ with $i = 1,2$ show 3D images obtained with our algorithm at the left ($i = 1$ for good reconstructions and $i = 2$ for worse reconstructions), the ground truth at the centre, and the SSIM maps between ground truth and reconstruction at the right. n3 (with n = a,b,c,d) are plots of the mean value of the SSIM maps for 500 different images. The analysis indicates that the algorithm performance remains relatively stable for each of the different type of objects.

\subsection{Amount of data required for training of the ANN}

\begin{figure*}
\centering
\includegraphics[width=0.60\linewidth]{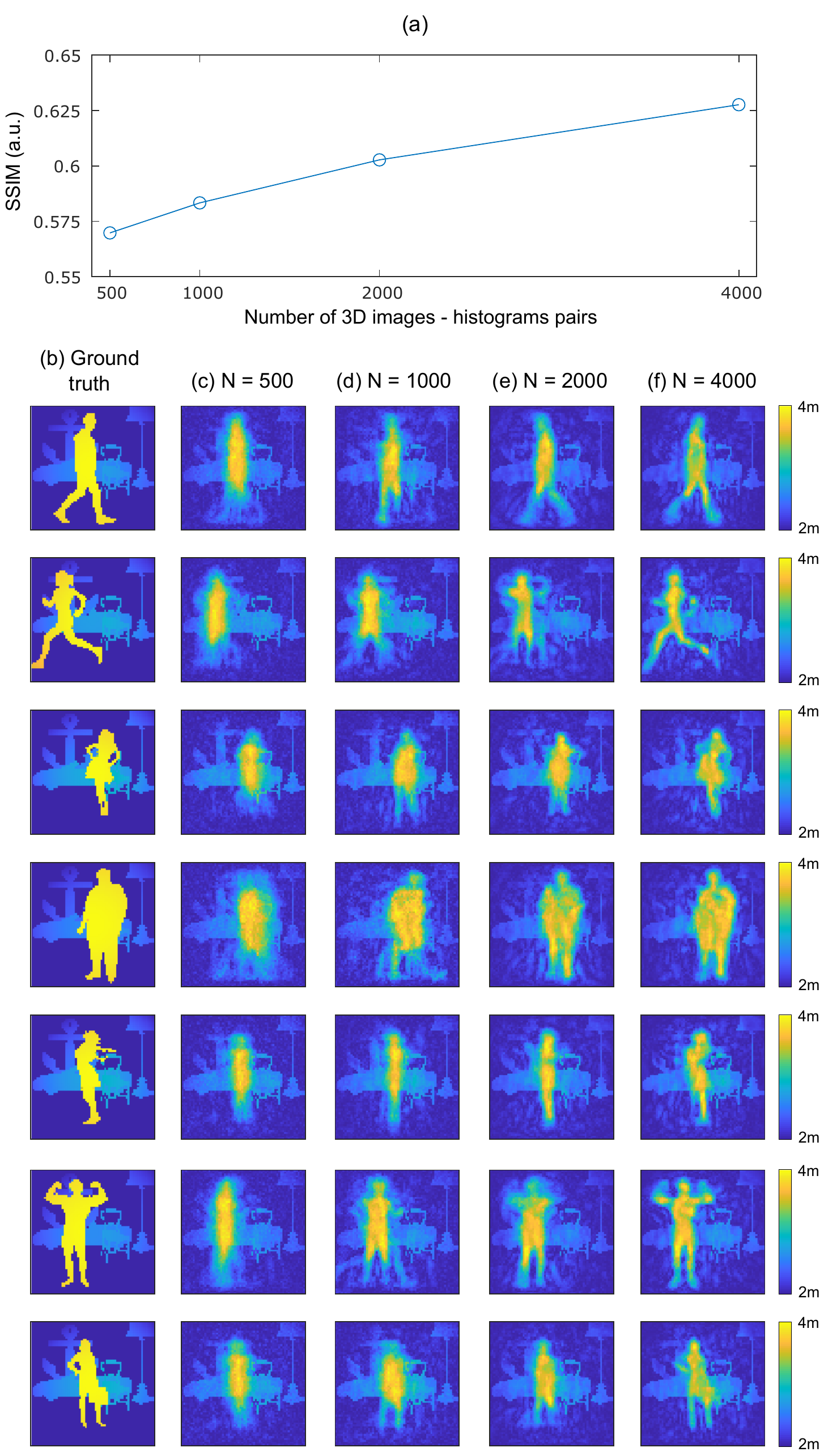}
\caption{
Reconstruction performance of the algorithm while increasing the data set size. (a) SSIM between test images and their corresponding reconstructions, obtained over 200 test pairs, for different sizes of the training data set. (b)-(f) examples of different test images obtained for training data sets with (c) $N = 500$, (d) $N = 1000$, (e) $N = 2000$, and (f) $N = 4000$ ToF images-histogram pairs, compared to their ground truths (b).
}
\label{sfdataset}
\end{figure*}

In machine learning techniques, one of key aspects to take into account is the size of the data set used for training of the algorithm. Very advanced algorithms tend to use tens of thousands (and even millions) of input-output pairs, but for our application we used only few thousands of samples. The final size of the data set was taken by inspection: we trained the algorithm multiple times with increasing number of 3D images-temporal histogram pairs until we achieved good enough reconstructions. 
Fig.~\ref{sfdataset} summarises the results of a numerical investigation, where we show the SSIM averaged over 200 test images for training data sets with dimensions $N = [ 500,1000,2000,4000]$, being $N = 4000$ the number used in the results shown throughout our work. Fig.~\ref{sfdataset}(a) shows the SSIM between test images and their corresponding reconstructions, obtained over 200 test pairs, for different sizes of the training data set. Figs.~\ref{sfdataset}(c)-(f) are examples of different test images obtained for training data sets with (c) $N = 500$, (d) $N = 1000$, (e) $N = 2000$, and (f) $N = 4000$ ToF images-histogram pairs, compared to their ground truths (b).
As expected, the more the data used, the more the reconstructed images resemble the ground truth. 
\subsection{Influence of noise}
\begin{figure*}
\centering
\includegraphics[width=0.75\linewidth]{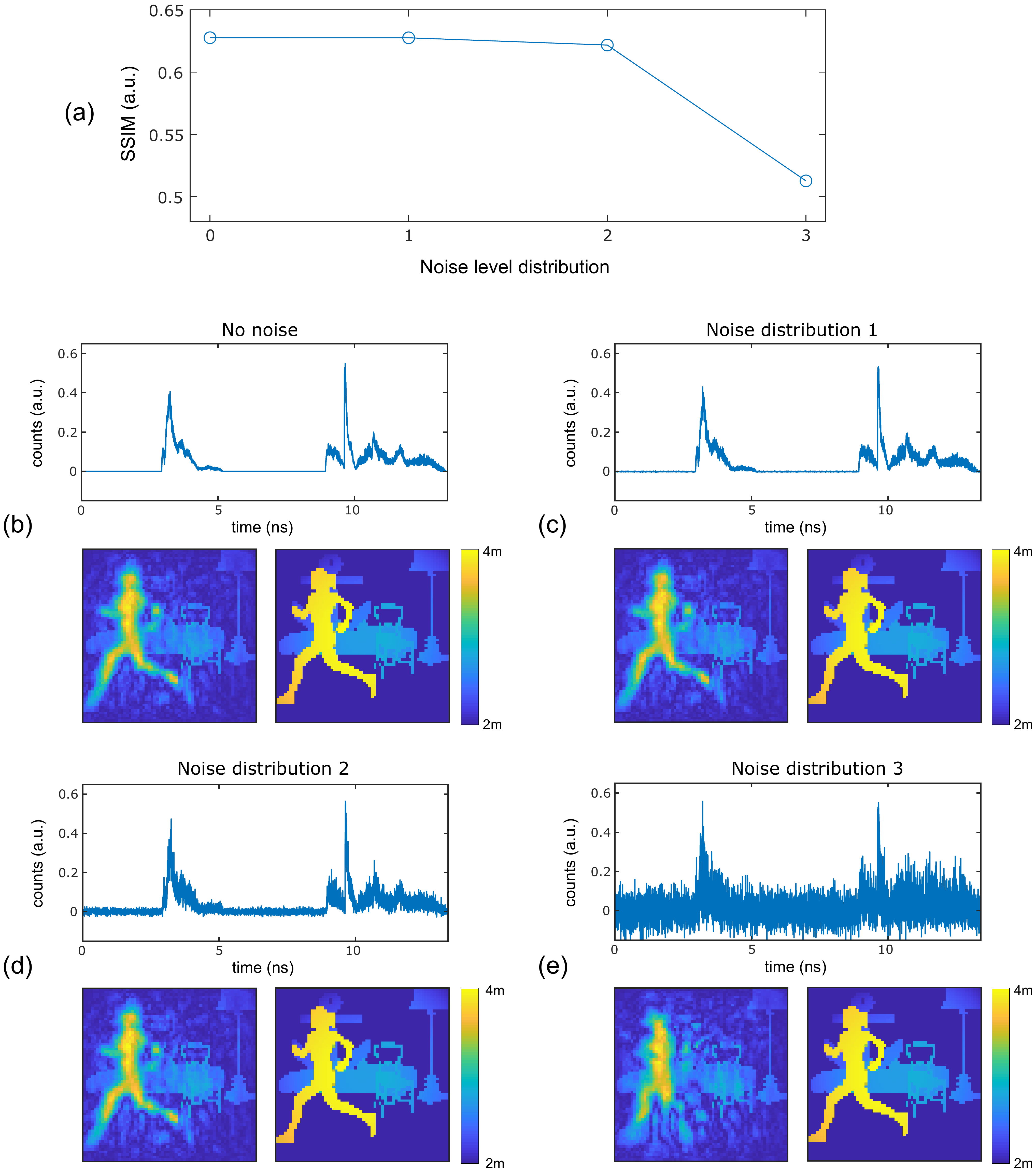}
\caption{
Performance of the reconstruction algorithm with noisy temporal histograms. (a) SSIM between test images and their corresponding reconstructions, obtained over 200 test pairs, for different noise distributions. `Noise level distribution 0' means no noise at all, while `Noise level distribution $i$' ($i = 1,2,3$) has added Poisson and Gaussian noise with noise expectation $\approx 3.2\%, 10\%, 33\%$, with respect to the temporal histogram signal. Each panel (b) - (e) give examples of histograms simulated from a single ToF with different noise distributions added (top figure), the ToF image these histograms were generated from (bottom right figure), and the corresponding images retrieved with our algorithm (bottom left figure). 
}
\label{sfnoise}
\end{figure*}

Here we show the resilience of the algorithm to noisy signals. To this aim, we numerically tested the reconstruction algorithm performance to signals with different amounts of noise. Note that the noise is directly added into the $\mathbf{x}$ term in Eq.~(\ref{perceptron}). The results are summarized in Fig.~\ref{sfnoise}: plot (a)
gives the SSIM between test images and their corresponding reconstructions, obtained over 200 test pairs, for different noise distributions. `Noise level distribution 0' means no noise at all, while `Noise level distribution $i$' ($i = 1,2,3$) has added Poisson and Gaussian noise with noise expectation $\approx 3.2\%, 10\%, 33\%$, with respect to the temporal histogram signal. Blocks (b) - (e) provide examples of histograms simulated from a single ToF with different noise distributions added (top figure), the ToF image these histograms were generated from (bottom right figure), and the corresponding images retrieved with our algorithm (bottom left figure). 
As it can be appreciated, our method can hold reasonable amounts of noise and provide images with sufficient quality in noisy environments. 

\subsection{Impact of non-uniform reflectivity}

\begin{figure*}
\centering
\includegraphics[width=0.60\linewidth]{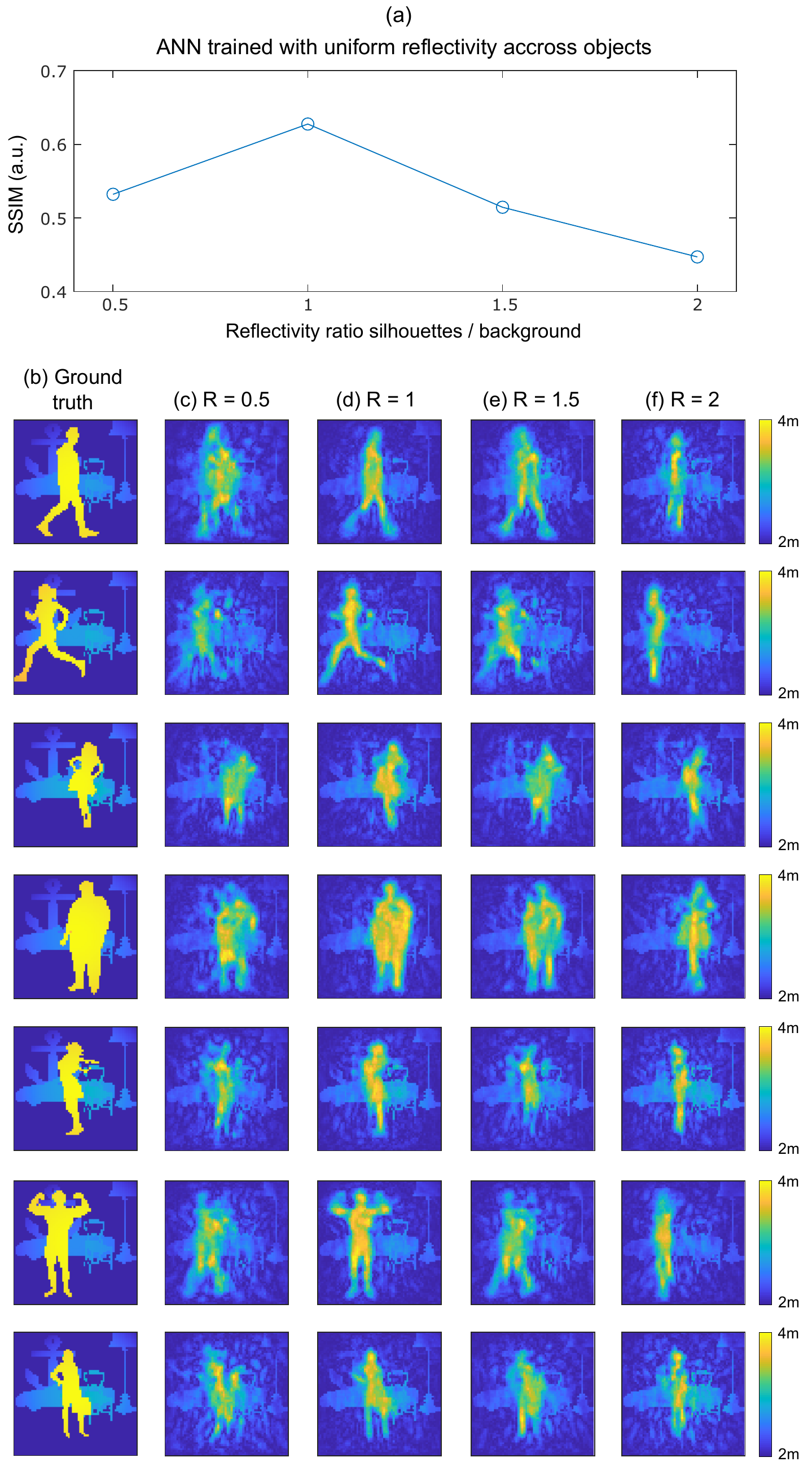}
\caption{
Reconstruction performance of the algorithm trained with uniform reflectivity across all objects while testing on objects with varying reflectivity. (a) SSIM between test images and their corresponding reconstructions, obtained over 200 test pairs, for different reflectivity ratios (R) between silhouettes and background. (b)-(f) examples of different test images obtained for training data sets with (c) $R = 0.5$, (d) $R = 1$, (e) $R = 1.5$, and (f) $R = 2$, compared to their ground truths (b).
}
\label{sfalbedoa}
\end{figure*}

For the data shown in the main document (both numerical and experimental), we used moving objects with uniform and constant reflectivity. However, there is an interesting question to explore, namely the fact that two identical objects with different reflectivity will manifest in the temporal histogram with peaks of different heights. To investigate this, we use numerical simulations to test the performance of our approach to retrieve images from scenes with non-uniform reflectivity. In our tests, we assume that the background objects of the scene have uniform reflectivity and we allow the reflectivity of the moving objects (the human silhouettes) to change. We quantify the reflectivity of the moving objects with respect to the background via the ratio $R = r_{\rm{silhouettes}}/r_{\rm{background}}$, where $r_{\rm{silhouettes}}$ and $r_{\rm{background}}$ are the reflectivity of the silhouettes and background, respectively. We first tested the reconstruction quality of the ANN when trained with scenes with uniform reflectivity (this is, with $R = 1$), and presented with histograms from scenes with reflectivity with values $R = {0.5, 1.0, 1.5, 2.0}$. Our results are summarized in  Fig.~\ref{sfalbedoa}. (a) shows the SSIM between test images and their corresponding reconstructions, obtained over 200 test pairs, for different reflectivity ratios (R) between silhouettes and background. (b)-(f) examples of different test images obtained for testing data sets with (c) $R = 0.5$, (d) $R = 1$, (e) $R = 1.5$, and (f) $R = 2$, compared to their ground truths (b). The results show that the algorithm struggles to retrieve correct images when $R \neq 1$, which can be seen as a potential limitation of the approach. 

\begin{figure*}
\centering
\includegraphics[width=0.60\linewidth]{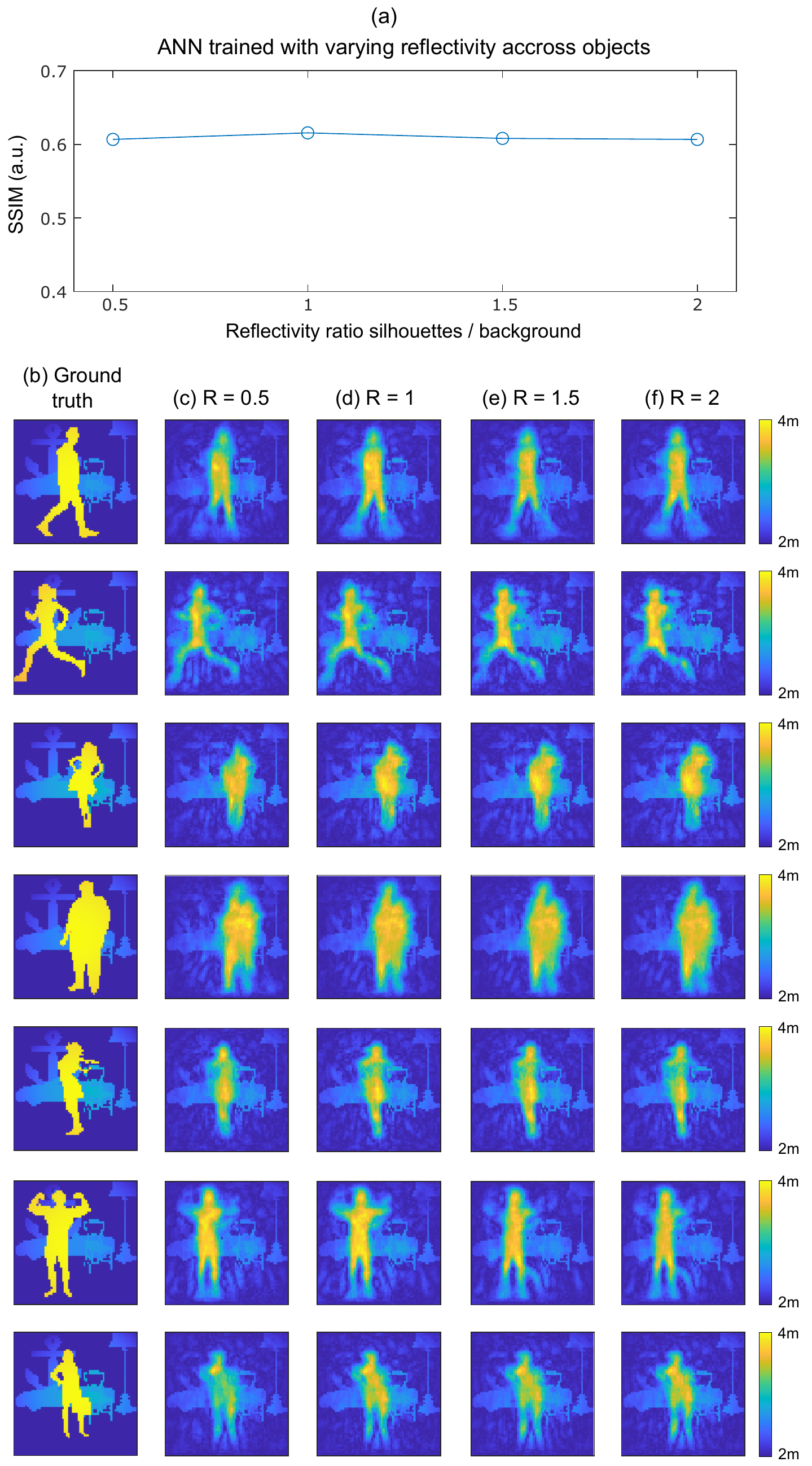}
\caption{
Reconstruction performance of the algorithm trained with varying reflectivity across objects. (a) SSIM between test images and their corresponding reconstructions, obtained over 200 test pairs, for different reflectivity ratios (R) between silhouettes and background. (b)-(f) examples of different test images obtained for training data sets with (c) $R = 0.5$, (d) $R = 1$, (e) $R = 1.5$, and (f) $R = 2$, compared to their ground truths (b).
}
\label{sfalbedob}
\end{figure*}

However, in order to have a fair analysis, we re-trained our algorithm with a data set containing moving objects with different reflectivity ratio in the range $R = [0.25,4.0]$. The results, see Fig.~\ref{sfalbedob}, indicate that in this case the algorithm is now more resilient to changes in the reflectivity of the objects and that it can retrieve images with similar SSIM even when $R = 2$. This example also shows that the data set used for training is key in order to retrieve good-quality images. Thus, by having a more complete and extended training data set, it seems possible to extend the applicability of our technique so as to correctly retrieve 3D images of objects with varying size and reflectivity.  

\subsection{Impact of the impulse response function}

\begin{figure*}[htb]
\centering
\includegraphics[width=0.85\linewidth]{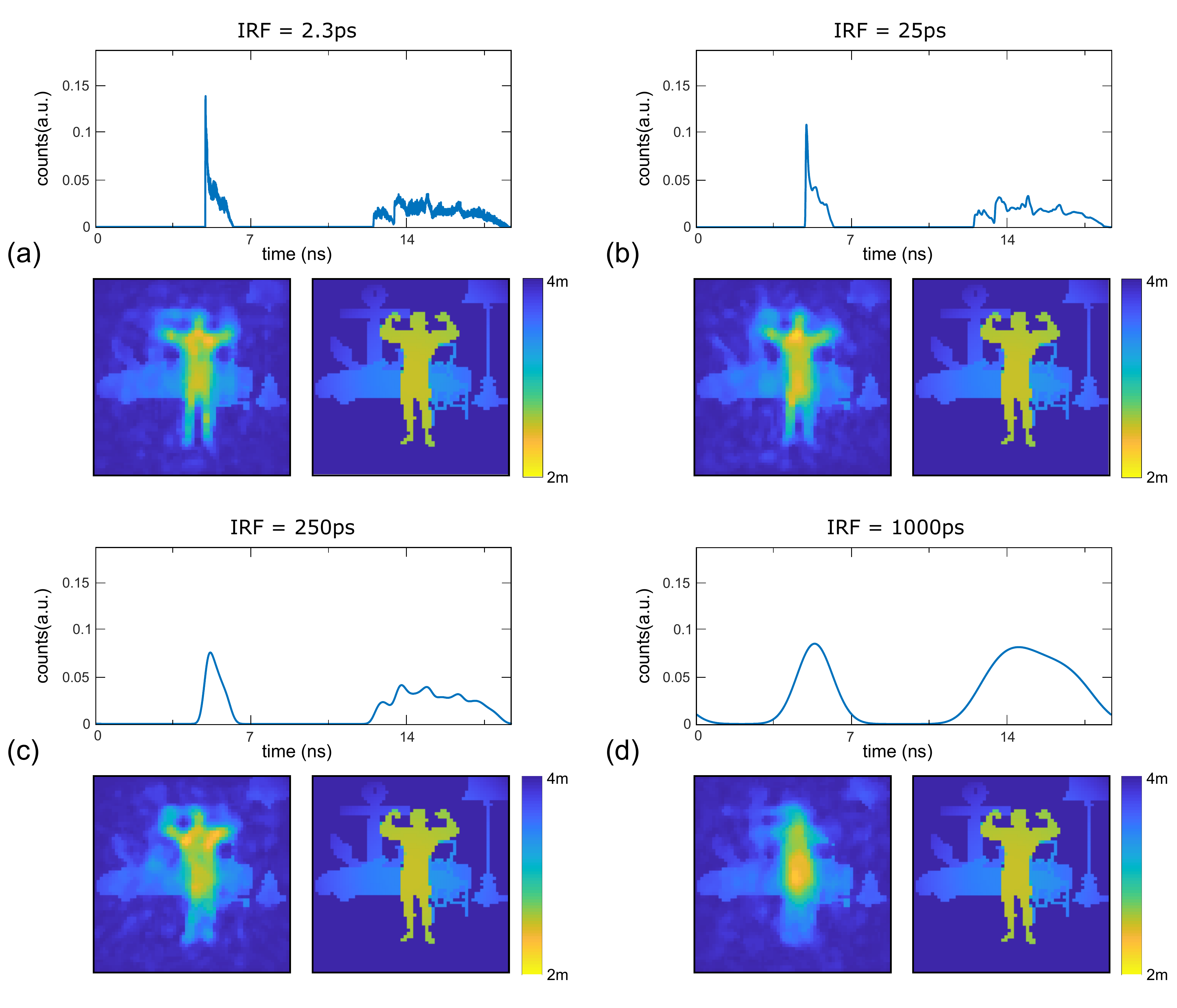}
\caption{
3D image reconstruction capabilities of our system when convolving the experimentally recorded temporal traces with Gaussian IRFs with: (a) $\Delta t = 2.3\,\rm{ps}$, (b) $\Delta t = 25\,\rm{ps}$, (c) $\Delta t = 250\,\rm{ps}$, and (d) $\Delta t = 1000\,\rm{ps}$. The top part in each image is the temporal histogram used to reconstruct the bottom-left image, while the bottom-right image is the ground ToF image for comparison.}
\label{sf6}
\end{figure*}

We also investigate how the reconstructed 3D images degrade as the IRF increases. Our starting point are the numerical simulations used above and in the main document, which correspond to the best-possible scenario achievable with state-of-art electronics providing $2.3\,\rm{ps}$ IRF. Any degradation of the system originated, for instance, from jitter, the use of longer laser pulses, and other related effects will manifest as a longer IRF. To account for this effect we take the best-possible-scenario histograms $h$ and convolve them with Gaussian IRFs $G$ with different widths $\Delta t$. This is strictly equivalent to change the IRF in our simulations but allows us to re-use the already generated histograms in the sections above. 
In other words, we replace $h$ with $\hat{h} = h \circledast G$, where $G = \exp{\left( -t^2 / \Delta t^2 \right)}$. 
The new histograms $\hat{h}$ are then paired with their corresponding ToF images and used to re-train the ANN. In Fig.~\ref{sf6} we summarize our results. Fig.~\ref{sf6} is divided in 4 blocks (a), (b), (c), (d) corresponding to results obtained for $\Delta t = [2.3,25,250,1000]\,\rm{ps}$, respectively. Therefore, block (a) corresponds to the best-case scenario. In each block, the top part shows the plot of corresponds to the temporal histogram used to reconstruct the bottom-left image, while the bottom-right image is the ground ToF image for comparison.
As expected, shorter IRFs improve the reconstruction of limbs and fine details of the objects. 

\subsection{Minimum resolvable transverse feature}

\begin{figure*}[htb]
\centering
\includegraphics[width=0.4\linewidth]{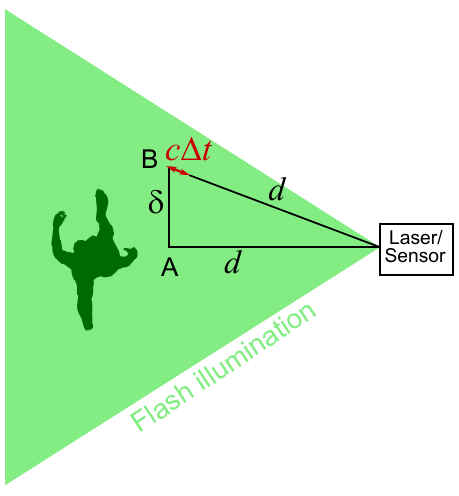}
\caption{
Geometry of single-point 3D imaging with time-resolving detectors. $\delta$ is the minimum resolvable feature obtained from a system with IRF $\Delta t$ at an axial distance $d$ from the detector. $c$ is the speed of light. The scene is seen from the top. 
}
\label{sf7}
\end{figure*}

We consider the situation depicted in Fig.~\ref{sf7}. The scene is flash illuminated with a pulsed laser and the back-scattered photons are collected by a single-point time-resolving sensor. We are interested to address the problem of resolving spatially two points $A$ and $B$ in space whose difference in time of flight is given by the IRF of the system, i.e $\Delta t$. These two points are separated by a transverse distance $\delta$ and we consider that one of them is co-axial with the laser and sensor. Given the geometry shown Fig.~\ref{sf7}, $A$ and $B$ form a triangle rectangle with the laser/sensor. As a consequence, by simple geometry, one obtains that:
\begin{equation}
    \delta = \sqrt{(d+ c\Delta t)^2 - d^2} = c \Delta t \sqrt{\frac{2d}{c \Delta t} + 1}.
    \label{eq1b}
\end{equation}

Eq.~(\ref{eq1a}) states that the minimum lateral resolvable distance $\delta$ depends not only on the IRF but also on the distance to the detector, following a square root law. 

\section{Cross-modality optical 3D imaging using a RADAR chip}

\begin{figure*}[htb]
\centering
\includegraphics[width=0.8\linewidth]{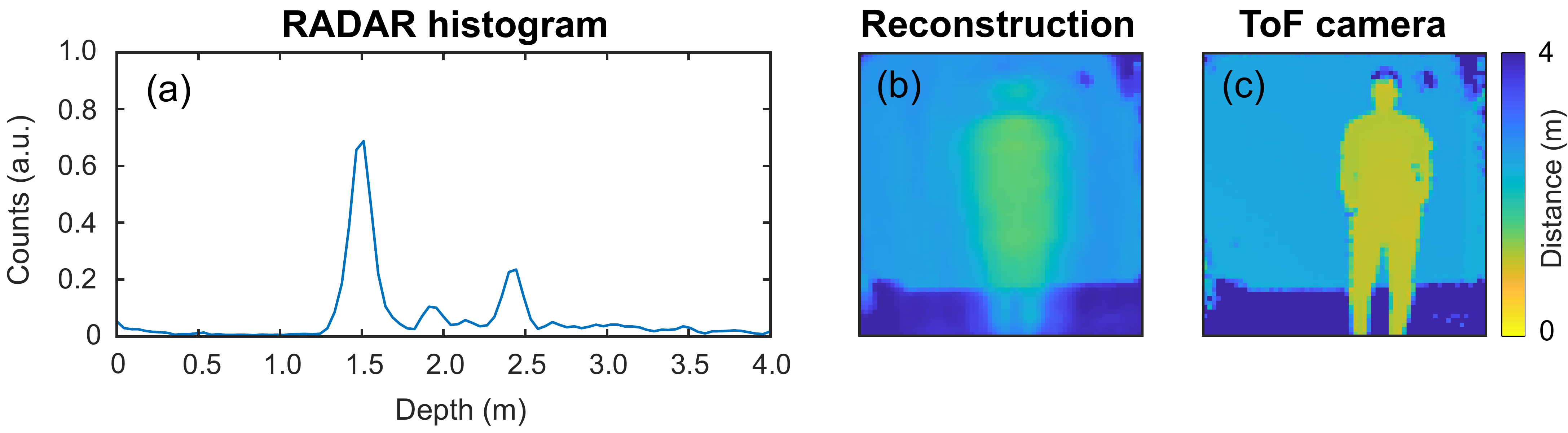}
\caption{
Experimental results showing cross-modality 3D imaging from a time histogram recorded with a impulse RADAR transceiver in the GHz regime. (a) Depicts the recorded temporal trace, (b) shows the reconstructed 3D scene, and (c) is the ToF depth-encoded image for comparison. A full video is available in
the supplementary information (Supplementary Video 2). The right-hand-side colour bar describes the colour-encoded depth map. }
\label{sf8}
\end{figure*}

Any sensing system that can map the position of objects within a scene into a temporal histogram can be used to obtain a 3D image from the scene. This opens new routes in imaging, for instance by using acoustic or radio waves. To test the feasibility of this concept, we performed new experiments with an impulse RADAR transceiver that emitted and collected pulsed electromagnetic radiation at $7.29\,\rm{GHz}$ (Novelda XeThru X4). The RADAR transceiver had a bandwidth of $1.4\,\rm{GHz}$ and a pulse duration of $670\,\rm{ps}$, which corresponds to a depth-spatial resolution of $c \tau = 20\,\rm{cm}$ and, according to the model presented above, to a transverse spatial resolution of $90 \,\rm{cm}$ at $2\,\rm{m}$ distance. Following the previously discussed procedure for the optical laser pulses, we gathered new data combining the impulse RADAR with the (optical) ToF camera and re-trained the ANN. In this case, only 3000 ToF images - temporal histogram pairs were used for training, which, together with the low temporal resolution of the RADAR chip, leads to a poorer image reconstruction compared to the one obtained with laser pulses (see Fig.~\ref{sf8}. However, we are still able to retrieve the general properties of the individual, e.g. their size, and position on the scene, see also Supplementary Video 2. 

\section{Pseudo-codes}

In this section we provide the pseudo-codes required to simulate the ToF data and to train the ANN. 

\subsection{ToF simulations}

{\fontfamily{pcr}\selectfont 
\noindent 1: Load file with multiple individuals $I_i$ with \\ \hspace*{0.6cm} $i = [1,10]$ \\
2: Define the range $R$ of depths and number \\ 
\hspace*{0.6cm} of positions in $(x,y,z)$ through which the  \\ 
\hspace*{0.6cm} individuals are positioned \\
3: \textbf{for} $I_i$ \textbf{do} \\
4: \hspace*{0.25cm} calculate coordinates $\rm{Coord}_i$ of $I_i$ \\
5: \hspace*{0.25cm} \textbf{for} all $(x_j,y_j,z_j)$ within $R$  \textbf{do} \\
6: \hspace*{0.75cm} $\rm{Coord}_i \xrightarrow{replace} \rm{Coord}_i(x_j,y_j,z_j)$ \\
7: \hspace*{0.75cm} scale $I_i$ by a factor $S = 2/z_j$ \\
8: \hspace*{0.75cm} create a new image: $\hat{I}_i = I_i(x_j,y_j,z_j)$   \\
\textcolor{white}{8:} \hspace*{0.75cm} scaled and displaced \\
9: \hspace*{0.25cm} \textbf{for} all pixels $p_k = (px_k,py_k)$ within $\hat{I}_i$  \textbf{do} \\
10: \hspace*{0.6cm} calculate the distance $d_k$ from sensor \\
\textcolor{white}{10:} \hspace*{0.6cm} [placed at $(x_0,y_0,z_0)$]: \\
\textcolor{white}{10:} \hspace*{0.75cm} $d_k=\sqrt{(x_k-x_0)^2+(y_k-y_0)^2+(z_k-z_0)^2}$ \\
11: \hspace*{0.60cm} calculate time of flight from \\ \textcolor{white}{12:} \hspace*{0.60cm} distance $d_k$: 
$t_k~=~ (2c)^{-1}d_k$ \\
12: \hspace*{0.60cm} estimate number of photons $n_k$ from \\ \textcolor{white}{12:} \hspace*{0.6cm} pixel $p_k$ according  to $d_k$: $n_k = P_0/d_k^4$\\
13: \hspace*{0.60cm} save $d_k,n_k$ into a variable $f_i,k = (d_k,n_k)$ \\
14: \hspace*{0.2cm} obtain histogram $h_i$ from $f_i$ \\
15: \hspace*{0.2cm} create ToF image by replacing value of \\
\textcolor{white}{12:} \hspace*{0.2cm} $p_k$ with $t_k$ \\
16: \hspace*{0.25cm} down-sample the ToF image to the de- \\
\textcolor{white}{12:} \hspace*{0.25cm} sired resolution for training the ANN 
}
\subsection{Artificial neural network}
{\fontfamily{pcr}\selectfont 
\noindent 1: Load ToF images $T_i$ and time histograms $h_i$ \\ \textcolor{white}{1:}  with $i = [1,N]$, $N$ number of images \\
2: Flatten every $T_i$ (with dimensions $d_1 \times d_2$) \\
\textcolor{white}{2:} into a single vector, with dimensions \\ 
\textcolor{white}{2:} $d = d_1*d_2$ \\
3: define architecture of the ANN model \textbf{M} \\
4: train(\textbf{M}) using $T$ and $h$ \\
5: save(\textbf{M}) \\
6: test(\textbf{M}) with unseen pairs of $T$ and $h$
}


\begin{thebibliography}{51}%
\makeatletter
\providecommand \@ifxundefined [1]{%
 \@ifx{#1\undefined}
}%
\providecommand \@ifnum [1]{%
 \ifnum #1\expandafter \@firstoftwo
 \else \expandafter \@secondoftwo
 \fi
}%
\providecommand \@ifx [1]{%
 \ifx #1\expandafter \@firstoftwo
 \else \expandafter \@secondoftwo
 \fi
}%
\providecommand \natexlab [1]{#1}%
\providecommand \enquote  [1]{``#1''}%
\providecommand \bibnamefont  [1]{#1}%
\providecommand \bibfnamefont [1]{#1}%
\providecommand \citenamefont [1]{#1}%
\providecommand \href@noop [0]{\@secondoftwo}%
\providecommand \href [0]{\begingroup \@sanitize@url \@href}%
\providecommand \@href[1]{\@@startlink{#1}\@@href}%
\providecommand \@@href[1]{\endgroup#1\@@endlink}%
\providecommand \@sanitize@url [0]{\catcode `\\12\catcode `\$12\catcode
  `\&12\catcode `\#12\catcode `\^12\catcode `\_12\catcode `\%12\relax}%
\providecommand \@@startlink[1]{}%
\providecommand \@@endlink[0]{}%
\providecommand \url  [0]{\begingroup\@sanitize@url \@url }%
\providecommand \@url [1]{\endgroup\@href {#1}{\urlprefix }}%
\providecommand \urlprefix  [0]{URL }%
\providecommand \Eprint [0]{\href }%
\providecommand \doibase [0]{http://dx.doi.org/}%
\providecommand \selectlanguage [0]{\@gobble}%
\providecommand \bibinfo  [0]{\@secondoftwo}%
\providecommand \bibfield  [0]{\@secondoftwo}%
\providecommand \translation [1]{[#1]}%
\providecommand \BibitemOpen [0]{}%
\providecommand \bibitemStop [0]{}%
\providecommand \bibitemNoStop [0]{.\EOS\space}%
\providecommand \EOS [0]{\spacefactor3000\relax}%
\providecommand \BibitemShut  [1]{\csname bibitem#1\endcsname}%
\let\auto@bib@innerbib\@empty
\bibitem [{\citenamefont {Shapiro}(2008)}]{2008:PRA:saphiro}%
  \BibitemOpen
  \bibfield  {author} {\bibinfo {author} {\bibfnamefont {J.~H.}\ \bibnamefont
  {Shapiro}},\ }\href@noop {} {\bibfield  {journal} {\bibinfo  {journal}
  {Physical Review A}\ }\textbf {\bibinfo {volume} {78}},\ \bibinfo {pages}
  {061802} (\bibinfo {year} {2008})}\BibitemShut {NoStop}%
\bibitem [{\citenamefont {Duarte}\ \emph {et~al.}(2008)\citenamefont {Duarte},
  \citenamefont {Davenport}, \citenamefont {Takhar}, \citenamefont {Laska},
  \citenamefont {Sun}, \citenamefont {Kelly},\ and\ \citenamefont
  {Baraniuk}}]{duarte:2008:IEEE}%
  \BibitemOpen
  \bibfield  {author} {\bibinfo {author} {\bibfnamefont {M.~F.}\ \bibnamefont
  {Duarte}}, \bibinfo {author} {\bibfnamefont {M.~A.}\ \bibnamefont
  {Davenport}}, \bibinfo {author} {\bibfnamefont {D.}~\bibnamefont {Takhar}},
  \bibinfo {author} {\bibfnamefont {J.~N.}\ \bibnamefont {Laska}}, \bibinfo
  {author} {\bibfnamefont {T.}~\bibnamefont {Sun}}, \bibinfo {author}
  {\bibfnamefont {K.~F.}\ \bibnamefont {Kelly}}, \ and\ \bibinfo {author}
  {\bibfnamefont {R.~G.}\ \bibnamefont {Baraniuk}},\ }\href@noop {} {\bibfield
  {journal} {\bibinfo  {journal} {IEEE signal processing magazine}\ }\textbf
  {\bibinfo {volume} {25}},\ \bibinfo {pages} {83} (\bibinfo {year}
  {2008})}\BibitemShut {NoStop}%
\bibitem [{\citenamefont {Edgar}\ \emph {et~al.}(2019)\citenamefont {Edgar},
  \citenamefont {Gibson},\ and\ \citenamefont
  {Padgett}}]{padgett:2019:natphoton}%
  \BibitemOpen
  \bibfield  {author} {\bibinfo {author} {\bibfnamefont {M.~P.}\ \bibnamefont
  {Edgar}}, \bibinfo {author} {\bibfnamefont {G.~M.}\ \bibnamefont {Gibson}}, \
  and\ \bibinfo {author} {\bibfnamefont {M.~J.}\ \bibnamefont {Padgett}},\
  }\href@noop {} {\bibfield  {journal} {\bibinfo  {journal} {Nature Photonics}\
  }\textbf {\bibinfo {volume} {13}},\ \bibinfo {pages} {13â€“20} (\bibinfo
  {year} {2019})}\BibitemShut {NoStop}%
\bibitem [{\citenamefont {Lindell}\ \emph {et~al.}(2018)\citenamefont
  {Lindell}, \citenamefont {O'Toole},\ and\ \citenamefont
  {Wetzstein}}]{wetzstein:2018:acm}%
  \BibitemOpen
  \bibfield  {author} {\bibinfo {author} {\bibfnamefont {D.~B.}\ \bibnamefont
  {Lindell}}, \bibinfo {author} {\bibfnamefont {M.}~\bibnamefont {O'Toole}}, \
  and\ \bibinfo {author} {\bibfnamefont {G.}~\bibnamefont {Wetzstein}},\
  }\href@noop {} {\bibfield  {journal} {\bibinfo  {journal} {ACM Trans.
  Graph.}\ }\textbf {\bibinfo {volume} {37}},\ \bibinfo {pages} {113} (\bibinfo
  {year} {2018})}\BibitemShut {NoStop}%
\bibitem [{\citenamefont {Radwell}\ \emph {et~al.}(2019)\citenamefont
  {Radwell}, \citenamefont {Selyem}, \citenamefont {Mertens}, \citenamefont
  {Edgar},\ and\ \citenamefont {Padgett}}]{radwell:2019:scirep}%
  \BibitemOpen
  \bibfield  {author} {\bibinfo {author} {\bibfnamefont {N.}~\bibnamefont
  {Radwell}}, \bibinfo {author} {\bibfnamefont {A.}~\bibnamefont {Selyem}},
  \bibinfo {author} {\bibfnamefont {L.}~\bibnamefont {Mertens}}, \bibinfo
  {author} {\bibfnamefont {M.~P.}\ \bibnamefont {Edgar}}, \ and\ \bibinfo
  {author} {\bibfnamefont {M.~J.}\ \bibnamefont {Padgett}},\ }\href@noop {}
  {\bibfield  {journal} {\bibinfo  {journal} {Scientific Reports}\ }\textbf
  {\bibinfo {volume} {9}},\ \bibinfo {pages} {5241} (\bibinfo {year}
  {2019})}\BibitemShut {NoStop}%
\bibitem [{\citenamefont {Callenberg}\ \emph {et~al.}(2019)\citenamefont
  {Callenberg}, \citenamefont {Lyons}, \citenamefont {den Brok}, \citenamefont
  {Henderson}, \citenamefont {Hullin},\ and\ \citenamefont
  {Faccio}}]{lyons:2019:CLEO}%
  \BibitemOpen
  \bibfield  {author} {\bibinfo {author} {\bibfnamefont {C.}~\bibnamefont
  {Callenberg}}, \bibinfo {author} {\bibfnamefont {A.}~\bibnamefont {Lyons}},
  \bibinfo {author} {\bibfnamefont {D.}~\bibnamefont {den Brok}}, \bibinfo
  {author} {\bibfnamefont {R.}~\bibnamefont {Henderson}}, \bibinfo {author}
  {\bibfnamefont {M.~B.}\ \bibnamefont {Hullin}}, \ and\ \bibinfo {author}
  {\bibfnamefont {D.}~\bibnamefont {Faccio}},\ }in\ \href@noop {} {\emph
  {\bibinfo {booktitle} {Computational Optical Sensing and Imaging}}}\
  (\bibinfo {organization} {Optical Society of America},\ \bibinfo {year}
  {2019})\ pp.\ \bibinfo {pages} {CTh2A--3}\BibitemShut {NoStop}%
\bibitem [{\citenamefont {Barnard}\ and\ \citenamefont
  {Fischler}(1982)}]{stereo}%
  \BibitemOpen
  \bibfield  {author} {\bibinfo {author} {\bibfnamefont {S.~T.}\ \bibnamefont
  {Barnard}}\ and\ \bibinfo {author} {\bibfnamefont {M.~A.}\ \bibnamefont
  {Fischler}},\ }\href {\doibase 10.1145/356893.356896} {\bibfield  {journal}
  {\bibinfo  {journal} {ACM Comput. Surv.}\ }\textbf {\bibinfo {volume} {14}},\
  \bibinfo {pages} {553} (\bibinfo {year} {1982})}\BibitemShut {NoStop}%
\bibitem [{\citenamefont {Frauel}\ \emph {et~al.}(2006)\citenamefont {Frauel},
  \citenamefont {Naughton}, \citenamefont {Matoba}, \citenamefont
  {Tajahuerce},\ and\ \citenamefont {Javidi}}]{holography}%
  \BibitemOpen
  \bibfield  {author} {\bibinfo {author} {\bibfnamefont {Y.}~\bibnamefont
  {Frauel}}, \bibinfo {author} {\bibfnamefont {T.~J.}\ \bibnamefont
  {Naughton}}, \bibinfo {author} {\bibfnamefont {O.}~\bibnamefont {Matoba}},
  \bibinfo {author} {\bibfnamefont {E.}~\bibnamefont {Tajahuerce}}, \ and\
  \bibinfo {author} {\bibfnamefont {B.}~\bibnamefont {Javidi}},\ }\href@noop {}
  {\bibfield  {journal} {\bibinfo  {journal} {Proceedings of the IEEE}\
  }\textbf {\bibinfo {volume} {94}},\ \bibinfo {pages} {636} (\bibinfo {year}
  {2006})}\BibitemShut {NoStop}%
\bibitem [{\citenamefont {Sun}\ \emph {et~al.}(2013)\citenamefont {Sun},
  \citenamefont {Edgar}, \citenamefont {Bowman}, \citenamefont {Vittert},
  \citenamefont {Welsh}, \citenamefont {Bowman},\ and\ \citenamefont
  {Padgett}}]{padgett:2013:science}%
  \BibitemOpen
  \bibfield  {author} {\bibinfo {author} {\bibfnamefont {B.}~\bibnamefont
  {Sun}}, \bibinfo {author} {\bibfnamefont {M.~P.}\ \bibnamefont {Edgar}},
  \bibinfo {author} {\bibfnamefont {R.}~\bibnamefont {Bowman}}, \bibinfo
  {author} {\bibfnamefont {L.~E.}\ \bibnamefont {Vittert}}, \bibinfo {author}
  {\bibfnamefont {S.}~\bibnamefont {Welsh}}, \bibinfo {author} {\bibfnamefont
  {A.}~\bibnamefont {Bowman}}, \ and\ \bibinfo {author} {\bibfnamefont {M.~J.}\
  \bibnamefont {Padgett}},\ }\href@noop {} {\bibfield  {journal} {\bibinfo
  {journal} {Science}\ }\textbf {\bibinfo {volume} {340}},\ \bibinfo {pages}
  {844} (\bibinfo {year} {2013})}\BibitemShut {NoStop}%
\bibitem [{\citenamefont {Sun}\ \emph {et~al.}(2016)\citenamefont {Sun},
  \citenamefont {Edgar}, \citenamefont {Gibson}, \citenamefont {Sun},
  \citenamefont {Radwell}, \citenamefont {Lamb},\ and\ \citenamefont
  {Padgett}}]{sun:2016:natcommun}%
  \BibitemOpen
  \bibfield  {author} {\bibinfo {author} {\bibfnamefont {M.-J.}\ \bibnamefont
  {Sun}}, \bibinfo {author} {\bibfnamefont {M.~P.}\ \bibnamefont {Edgar}},
  \bibinfo {author} {\bibfnamefont {G.~M.}\ \bibnamefont {Gibson}}, \bibinfo
  {author} {\bibfnamefont {B.}~\bibnamefont {Sun}}, \bibinfo {author}
  {\bibfnamefont {N.}~\bibnamefont {Radwell}}, \bibinfo {author} {\bibfnamefont
  {R.}~\bibnamefont {Lamb}}, \ and\ \bibinfo {author} {\bibfnamefont {M.~J.}\
  \bibnamefont {Padgett}},\ }\href@noop {} {\bibfield  {journal} {\bibinfo
  {journal} {Nature Communications}\ }\textbf {\bibinfo {volume} {7}},\
  \bibinfo {pages} {12010} (\bibinfo {year} {2016})}\BibitemShut {NoStop}%
\bibitem [{\citenamefont {Dong}\ and\ \citenamefont {Chen}(2017)}]{book-lidar}%
  \BibitemOpen
  \bibfield  {author} {\bibinfo {author} {\bibfnamefont {P.}~\bibnamefont
  {Dong}}\ and\ \bibinfo {author} {\bibfnamefont {Q.}~\bibnamefont {Chen}},\
  }\href@noop {} {\emph {\bibinfo {title} {LiDAR Remote Sensing and
  Applications}}}\ (\bibinfo  {publisher} {CRC Press},\ \bibinfo {year}
  {2017})\BibitemShut {NoStop}%
\bibitem [{\citenamefont {Kirmani}\ \emph {et~al.}(2014)\citenamefont
  {Kirmani}, \citenamefont {Venkatraman}, \citenamefont {Shin}, \citenamefont
  {Colaco}, \citenamefont {Wong}, \citenamefont {Shapiro1},\ and\ \citenamefont
  {Goyal}}]{first_photon}%
  \BibitemOpen
  \bibfield  {author} {\bibinfo {author} {\bibfnamefont {A.}~\bibnamefont
  {Kirmani}}, \bibinfo {author} {\bibfnamefont {D.}~\bibnamefont
  {Venkatraman}}, \bibinfo {author} {\bibfnamefont {D.}~\bibnamefont {Shin}},
  \bibinfo {author} {\bibfnamefont {A.}~\bibnamefont {Colaco}}, \bibinfo
  {author} {\bibfnamefont {F.~N.~C.}\ \bibnamefont {Wong}}, \bibinfo {author}
  {\bibfnamefont {J.~H.}\ \bibnamefont {Shapiro1}}, \ and\ \bibinfo {author}
  {\bibfnamefont {V.~K.}\ \bibnamefont {Goyal}},\ }\href@noop {} {\bibfield
  {journal} {\bibinfo  {journal} {Science}\ }\textbf {\bibinfo {volume}
  {343}},\ \bibinfo {pages} {58} (\bibinfo {year} {2014})}\BibitemShut
  {NoStop}%
\bibitem [{\citenamefont {Morris}\ \emph {et~al.}(2015)\citenamefont {Morris},
  \citenamefont {Aspden}, \citenamefont {Bell}, \citenamefont {Boyd},\ and\
  \citenamefont {Padgett}}]{morris}%
  \BibitemOpen
  \bibfield  {author} {\bibinfo {author} {\bibfnamefont {P.~A.}\ \bibnamefont
  {Morris}}, \bibinfo {author} {\bibfnamefont {R.~S.}\ \bibnamefont {Aspden}},
  \bibinfo {author} {\bibfnamefont {J.~E.~C.}\ \bibnamefont {Bell}}, \bibinfo
  {author} {\bibfnamefont {R.~W.}\ \bibnamefont {Boyd}}, \ and\ \bibinfo
  {author} {\bibfnamefont {M.~J.}\ \bibnamefont {Padgett}},\ }\href@noop {}
  {\bibfield  {journal} {\bibinfo  {journal} {Nat. Commun.}\ }\textbf {\bibinfo
  {volume} {6}},\ \bibinfo {pages} {5913} (\bibinfo {year} {2015})}\BibitemShut
  {NoStop}%
\bibitem [{\citenamefont {Tachella}\ \emph {et~al.}(2019)\citenamefont
  {Tachella}, \citenamefont {Altmann}, \citenamefont {Mellado}, \citenamefont
  {McCarthy}, \citenamefont {Tobin}, \citenamefont {Gerald S.~Buller},\ and\
  \citenamefont {McLaughlin}}]{altmann}%
  \BibitemOpen
  \bibfield  {author} {\bibinfo {author} {\bibfnamefont {J.}~\bibnamefont
  {Tachella}}, \bibinfo {author} {\bibfnamefont {Y.}~\bibnamefont {Altmann}},
  \bibinfo {author} {\bibfnamefont {N.}~\bibnamefont {Mellado}}, \bibinfo
  {author} {\bibfnamefont {A.}~\bibnamefont {McCarthy}}, \bibinfo {author}
  {\bibfnamefont {R.}~\bibnamefont {Tobin}}, \bibinfo {author} {\bibfnamefont
  {J.-Y.~T.}\ \bibnamefont {Gerald S.~Buller}}, \ and\ \bibinfo {author}
  {\bibfnamefont {S.}~\bibnamefont {McLaughlin}},\ }\href@noop {} {\bibfield
  {journal} {\bibinfo  {journal} {Nat. Commun.}\ }\textbf {\bibinfo {volume}
  {10}},\ \bibinfo {pages} {4984} (\bibinfo {year} {2019})}\BibitemShut
  {NoStop}%
\bibitem [{\citenamefont {Velten}\ \emph {et~al.}(2012)\citenamefont {Velten},
  \citenamefont {Willwacher}, \citenamefont {Gupta}, \citenamefont
  {Veeraraghavan}, \citenamefont {Bawendi},\ and\ \citenamefont
  {Raskar}}]{velten:2012:natcommun}%
  \BibitemOpen
  \bibfield  {author} {\bibinfo {author} {\bibfnamefont {A.}~\bibnamefont
  {Velten}}, \bibinfo {author} {\bibfnamefont {T.}~\bibnamefont {Willwacher}},
  \bibinfo {author} {\bibfnamefont {O.}~\bibnamefont {Gupta}}, \bibinfo
  {author} {\bibfnamefont {A.}~\bibnamefont {Veeraraghavan}}, \bibinfo {author}
  {\bibfnamefont {M.~G.}\ \bibnamefont {Bawendi}}, \ and\ \bibinfo {author}
  {\bibfnamefont {R.}~\bibnamefont {Raskar}},\ }\href@noop {} {\bibfield
  {journal} {\bibinfo  {journal} {Nature Communications}\ }\textbf {\bibinfo
  {volume} {3}},\ \bibinfo {pages} {745} (\bibinfo {year} {2012})}\BibitemShut
  {NoStop}%
\bibitem [{\citenamefont {Gariepy}\ \emph
  {et~al.}(2016{\natexlab{a}})\citenamefont {Gariepy}, \citenamefont
  {Tonolini}, \citenamefont {Henderson}, \citenamefont {Leach},\ and\
  \citenamefont {Faccio}}]{gariepy:2016:natphoton}%
  \BibitemOpen
  \bibfield  {author} {\bibinfo {author} {\bibfnamefont {G.}~\bibnamefont
  {Gariepy}}, \bibinfo {author} {\bibfnamefont {F.}~\bibnamefont {Tonolini}},
  \bibinfo {author} {\bibfnamefont {R.}~\bibnamefont {Henderson}}, \bibinfo
  {author} {\bibfnamefont {J.}~\bibnamefont {Leach}}, \ and\ \bibinfo {author}
  {\bibfnamefont {D.}~\bibnamefont {Faccio}},\ }\href@noop {} {\bibfield
  {journal} {\bibinfo  {journal} {Nature Photonics}\ }\textbf {\bibinfo
  {volume} {10}},\ \bibinfo {pages} {23} (\bibinfo {year}
  {2016}{\natexlab{a}})}\BibitemShut {NoStop}%
\bibitem [{\citenamefont {O'Toole}\ \emph {et~al.}(2018)\citenamefont
  {O'Toole}, \citenamefont {Lindell},\ and\ \citenamefont
  {Wetzstein}}]{wetzstein:2018:nature}%
  \BibitemOpen
  \bibfield  {author} {\bibinfo {author} {\bibfnamefont {M.}~\bibnamefont
  {O'Toole}}, \bibinfo {author} {\bibfnamefont {D.~B.}\ \bibnamefont
  {Lindell}}, \ and\ \bibinfo {author} {\bibfnamefont {G.}~\bibnamefont
  {Wetzstein}},\ }\href@noop {} {\bibfield  {journal} {\bibinfo  {journal}
  {Nature}\ }\textbf {\bibinfo {volume} {555}},\ \bibinfo {pages} {338â€“341}
  (\bibinfo {year} {2018})}\BibitemShut {NoStop}%
\bibitem [{\citenamefont {Jin}\ \emph {et~al.}(2018)\citenamefont {Jin},
  \citenamefont {Xie}, \citenamefont {Zhang}, \citenamefont {Zhang},\ and\
  \citenamefont {Zhao}}]{jin:2018:oe}%
  \BibitemOpen
  \bibfield  {author} {\bibinfo {author} {\bibfnamefont {C.}~\bibnamefont
  {Jin}}, \bibinfo {author} {\bibfnamefont {J.}~\bibnamefont {Xie}}, \bibinfo
  {author} {\bibfnamefont {S.}~\bibnamefont {Zhang}}, \bibinfo {author}
  {\bibfnamefont {Z.}~\bibnamefont {Zhang}}, \ and\ \bibinfo {author}
  {\bibfnamefont {Y.}~\bibnamefont {Zhao}},\ }\href@noop {} {\bibfield
  {journal} {\bibinfo  {journal} {Optics Express}\ }\textbf {\bibinfo {volume}
  {26}},\ \bibinfo {pages} {20089} (\bibinfo {year} {2018})}\BibitemShut
  {NoStop}%
\bibitem [{\citenamefont {Arellano}\ \emph {et~al.}(2017)\citenamefont
  {Arellano}, \citenamefont {Gutierrez},\ and\ \citenamefont
  {Jarabo}}]{jarabo:2017:oe}%
  \BibitemOpen
  \bibfield  {author} {\bibinfo {author} {\bibfnamefont {V.}~\bibnamefont
  {Arellano}}, \bibinfo {author} {\bibfnamefont {D.}~\bibnamefont {Gutierrez}},
  \ and\ \bibinfo {author} {\bibfnamefont {A.}~\bibnamefont {Jarabo}},\
  }\href@noop {} {\bibfield  {journal} {\bibinfo  {journal} {Optics Express}\
  }\textbf {\bibinfo {volume} {25}},\ \bibinfo {pages} {11574} (\bibinfo {year}
  {2017})}\BibitemShut {NoStop}%
\bibitem [{\citenamefont {Musarra}\ \emph {et~al.}(2019)\citenamefont
  {Musarra}, \citenamefont {Lyons}, \citenamefont {Conca}, \citenamefont
  {Altmann}, \citenamefont {Villa}, \citenamefont {Zappa}, \citenamefont
  {Padgett},\ and\ \citenamefont {Faccio}}]{musarra:2019:arxiv}%
  \BibitemOpen
  \bibfield  {author} {\bibinfo {author} {\bibfnamefont {G.}~\bibnamefont
  {Musarra}}, \bibinfo {author} {\bibfnamefont {A.}~\bibnamefont {Lyons}},
  \bibinfo {author} {\bibfnamefont {E.}~\bibnamefont {Conca}}, \bibinfo
  {author} {\bibfnamefont {Y.}~\bibnamefont {Altmann}}, \bibinfo {author}
  {\bibfnamefont {F.}~\bibnamefont {Villa}}, \bibinfo {author} {\bibfnamefont
  {F.}~\bibnamefont {Zappa}}, \bibinfo {author} {\bibfnamefont
  {M.}~\bibnamefont {Padgett}}, \ and\ \bibinfo {author} {\bibfnamefont
  {D.}~\bibnamefont {Faccio}},\ }\href@noop {} {\bibfield  {journal} {\bibinfo
  {journal} {Physical Review Applied}\ }\textbf {\bibinfo {volume} {12}},\
  \bibinfo {pages} {011002} (\bibinfo {year} {2019})}\BibitemShut {NoStop}%
\bibitem [{\citenamefont {Altmann}\ \emph {et~al.}(2018)\citenamefont
  {Altmann}, \citenamefont {McLaughlin}, \citenamefont {Padgett}, \citenamefont
  {Goyal}, \citenamefont {Hero},\ and\ \citenamefont
  {Faccio}}]{faccio:2018:science}%
  \BibitemOpen
  \bibfield  {author} {\bibinfo {author} {\bibfnamefont {Y.}~\bibnamefont
  {Altmann}}, \bibinfo {author} {\bibfnamefont {S.}~\bibnamefont {McLaughlin}},
  \bibinfo {author} {\bibfnamefont {M.~J.}\ \bibnamefont {Padgett}}, \bibinfo
  {author} {\bibfnamefont {V.~K.}\ \bibnamefont {Goyal}}, \bibinfo {author}
  {\bibfnamefont {A.~O.}\ \bibnamefont {Hero}}, \ and\ \bibinfo {author}
  {\bibfnamefont {D.}~\bibnamefont {Faccio}},\ }\href@noop {} {\bibfield
  {journal} {\bibinfo  {journal} {Science}\ }\textbf {\bibinfo {volume}
  {361}},\ \bibinfo {pages} {660} (\bibinfo {year} {2018})}\BibitemShut
  {NoStop}%
\bibitem [{\citenamefont {Jordan}\ and\ \citenamefont
  {Mitchell}(2015)}]{jordan:2015:nature}%
  \BibitemOpen
  \bibfield  {author} {\bibinfo {author} {\bibfnamefont {M.~I.}\ \bibnamefont
  {Jordan}}\ and\ \bibinfo {author} {\bibfnamefont {T.~M.}\ \bibnamefont
  {Mitchell}},\ }\href@noop {} {\bibfield  {journal} {\bibinfo  {journal}
  {Science}\ }\textbf {\bibinfo {volume} {349}},\ \bibinfo {pages} {255}
  (\bibinfo {year} {2015})}\BibitemShut {NoStop}%
\bibitem [{\citenamefont {LeCun}\ \emph {et~al.}(2015)\citenamefont {LeCun},
  \citenamefont {Bengio},\ and\ \citenamefont {Hinton}}]{lecun:2015:nature}%
  \BibitemOpen
  \bibfield  {author} {\bibinfo {author} {\bibfnamefont {Y.}~\bibnamefont
  {LeCun}}, \bibinfo {author} {\bibfnamefont {Y.}~\bibnamefont {Bengio}}, \
  and\ \bibinfo {author} {\bibfnamefont {G.}~\bibnamefont {Hinton}},\
  }\href@noop {} {\bibfield  {journal} {\bibinfo  {journal} {Nature}\ }\textbf
  {\bibinfo {volume} {521}},\ \bibinfo {pages} {436} (\bibinfo {year}
  {2015})}\BibitemShut {NoStop}%
\bibitem [{\citenamefont {Barbastathis}\ \emph {et~al.}(2019)\citenamefont
  {Barbastathis}, \citenamefont {Ozcan},\ and\ \citenamefont
  {Situ}}]{2019:ozcan:optica}%
  \BibitemOpen
  \bibfield  {author} {\bibinfo {author} {\bibfnamefont {G.}~\bibnamefont
  {Barbastathis}}, \bibinfo {author} {\bibfnamefont {A.}~\bibnamefont {Ozcan}},
  \ and\ \bibinfo {author} {\bibfnamefont {G.}~\bibnamefont {Situ}},\ }\href
  {\doibase 10.1364/OPTICA.6.000921} {\bibfield  {journal} {\bibinfo  {journal}
  {Optica}\ }\textbf {\bibinfo {volume} {6}},\ \bibinfo {pages} {921} (\bibinfo
  {year} {2019})}\BibitemShut {NoStop}%
\bibitem [{\citenamefont {Waller}\ and\ \citenamefont
  {Tian}(2015)}]{waller:2015:nature}%
  \BibitemOpen
  \bibfield  {author} {\bibinfo {author} {\bibfnamefont {L.}~\bibnamefont
  {Waller}}\ and\ \bibinfo {author} {\bibfnamefont {L.}~\bibnamefont {Tian}},\
  }\href@noop {} {\bibfield  {journal} {\bibinfo  {journal} {Nature}\ }\textbf
  {\bibinfo {volume} {523}},\ \bibinfo {pages} {416} (\bibinfo {year}
  {2015})}\BibitemShut {NoStop}%
\bibitem [{\citenamefont {Rivenson}\ \emph {et~al.}(2017)\citenamefont
  {Rivenson}, \citenamefont {GÃ¶rÃ¶cs}, \citenamefont {GÃ¼naydin},
  \citenamefont {Zhang}, \citenamefont {Wang},\ and\ \citenamefont
  {Ozcan}}]{ozcan:2017:optica}%
  \BibitemOpen
  \bibfield  {author} {\bibinfo {author} {\bibfnamefont {Y.}~\bibnamefont
  {Rivenson}}, \bibinfo {author} {\bibfnamefont {Z.}~\bibnamefont {Garacs}},
  \bibinfo {author} {\bibfnamefont {H.}~\bibnamefont {GÃ¼naydin}}, \bibinfo
  {author} {\bibfnamefont {Y.}~\bibnamefont {Zhang}}, \bibinfo {author}
  {\bibfnamefont {H.}~\bibnamefont {Wang}}, \ and\ \bibinfo {author}
  {\bibfnamefont {A.}~\bibnamefont {Ozcan}},\ }\href@noop {} {\bibfield
  {journal} {\bibinfo  {journal} {Optica}\ }\textbf {\bibinfo {volume} {4}},\
  \bibinfo {pages} {1437} (\bibinfo {year} {2017})}\BibitemShut {NoStop}%
\bibitem [{\citenamefont {Nehme}\ \emph {et~al.}(2018)\citenamefont {Nehme},
  \citenamefont {Weiss}, \citenamefont {Michaeli},\ and\ \citenamefont
  {Shechtman}}]{schechtman:2018:optica}%
  \BibitemOpen
  \bibfield  {author} {\bibinfo {author} {\bibfnamefont {E.}~\bibnamefont
  {Nehme}}, \bibinfo {author} {\bibfnamefont {L.~E.}\ \bibnamefont {Weiss}},
  \bibinfo {author} {\bibfnamefont {T.}~\bibnamefont {Michaeli}}, \ and\
  \bibinfo {author} {\bibfnamefont {Y.}~\bibnamefont {Shechtman}},\ }\href@noop
  {} {\bibfield  {journal} {\bibinfo  {journal} {Optica}\ }\textbf {\bibinfo
  {volume} {5}},\ \bibinfo {pages} {458} (\bibinfo {year} {2018})}\BibitemShut
  {NoStop}%
\bibitem [{\citenamefont {Wang}\ \emph {et~al.}(2019)\citenamefont {Wang},
  \citenamefont {Rivenson}, \citenamefont {Jin}, \citenamefont {Wei},
  \citenamefont {Gao}, \citenamefont {GÃ¼nayd{\i}n}, \citenamefont {Bentolila},
  \citenamefont {Kural},\ and\ \citenamefont {Ozcan}}]{ozcan:2019:natmeth}%
  \BibitemOpen
  \bibfield  {author} {\bibinfo {author} {\bibfnamefont {H.}~\bibnamefont
  {Wang}}, \bibinfo {author} {\bibfnamefont {Y.}~\bibnamefont {Rivenson}},
  \bibinfo {author} {\bibfnamefont {Y.}~\bibnamefont {Jin}}, \bibinfo {author}
  {\bibfnamefont {Z.}~\bibnamefont {Wei}}, \bibinfo {author} {\bibfnamefont
  {R.}~\bibnamefont {Gao}}, \bibinfo {author} {\bibfnamefont {H.}~\bibnamefont
  {Ganayd{\i}n}}, \bibinfo {author} {\bibfnamefont {L.~A.}\ \bibnamefont
  {Bentolila}}, \bibinfo {author} {\bibfnamefont {C.}~\bibnamefont {Kural}}, \
  and\ \bibinfo {author} {\bibfnamefont {A.}~\bibnamefont {Ozcan}},\
  }\href@noop {} {\bibfield  {journal} {\bibinfo  {journal} {Nature Methods}\
  }\textbf {\bibinfo {volume} {16}},\ \bibinfo {pages} {103} (\bibinfo {year}
  {2019})}\BibitemShut {NoStop}%
\bibitem [{\citenamefont {Rivenson}\ \emph
  {et~al.}(2018{\natexlab{a}})\citenamefont {Rivenson}, \citenamefont {Zhang},
  \citenamefont {GÃ¼nayd{\i}n}, \citenamefont {Teng},\ and\ \citenamefont
  {Ozcan}}]{ozcan:2018:lsa}%
  \BibitemOpen
  \bibfield  {author} {\bibinfo {author} {\bibfnamefont {Y.}~\bibnamefont
  {Rivenson}}, \bibinfo {author} {\bibfnamefont {Y.}~\bibnamefont {Zhang}},
  \bibinfo {author} {\bibfnamefont {H.}~\bibnamefont {GÃ¼nayd{\i}n}}, \bibinfo
  {author} {\bibfnamefont {D.}~\bibnamefont {Teng}}, \ and\ \bibinfo {author}
  {\bibfnamefont {A.}~\bibnamefont {Ozcan}},\ }\href@noop {} {\bibfield
  {journal} {\bibinfo  {journal} {Light: Science \& Applications}\ }\textbf
  {\bibinfo {volume} {7}},\ \bibinfo {pages} {17141} (\bibinfo {year}
  {2018}{\natexlab{a}})}\BibitemShut {NoStop}%
\bibitem [{\citenamefont {Goy}\ \emph {et~al.}(2018)\citenamefont {Goy},
  \citenamefont {Arthur}, \citenamefont {Li},\ and\ \citenamefont
  {Barbastathis}}]{barbastathis:2018:prl}%
  \BibitemOpen
  \bibfield  {author} {\bibinfo {author} {\bibfnamefont {A.}~\bibnamefont
  {Goy}}, \bibinfo {author} {\bibfnamefont {K.}~\bibnamefont {Arthur}},
  \bibinfo {author} {\bibfnamefont {S.}~\bibnamefont {Li}}, \ and\ \bibinfo
  {author} {\bibfnamefont {G.}~\bibnamefont {Barbastathis}},\ }\href@noop {}
  {\bibfield  {journal} {\bibinfo  {journal} {Physical Review Letters}\
  }\textbf {\bibinfo {volume} {121}},\ \bibinfo {pages} {243902} (\bibinfo
  {year} {2018})}\BibitemShut {NoStop}%
\bibitem [{\citenamefont {Sinha}\ \emph {et~al.}(2017)\citenamefont {Sinha},
  \citenamefont {Lee}, \citenamefont {Li},\ and\ \citenamefont
  {Barbastathis}}]{barbastathis:2017:optica}%
  \BibitemOpen
  \bibfield  {author} {\bibinfo {author} {\bibfnamefont {A.}~\bibnamefont
  {Sinha}}, \bibinfo {author} {\bibfnamefont {J.}~\bibnamefont {Lee}}, \bibinfo
  {author} {\bibfnamefont {S.}~\bibnamefont {Li}}, \ and\ \bibinfo {author}
  {\bibfnamefont {G.}~\bibnamefont {Barbastathis}},\ }\href@noop {} {\bibfield
  {journal} {\bibinfo  {journal} {Optica}\ }\textbf {\bibinfo {volume} {4}},\
  \bibinfo {pages} {1117} (\bibinfo {year} {2017})}\BibitemShut {NoStop}%
\bibitem [{\citenamefont {Rivenson}\ \emph
  {et~al.}(2018{\natexlab{b}})\citenamefont {Rivenson}, \citenamefont
  {Ceylan~Koydemir}, \citenamefont {Wang}, \citenamefont {Wei}, \citenamefont
  {Ren}, \citenamefont {Gunayd{\i}n}, \citenamefont {Zhang}, \citenamefont
  {Gorocs}, \citenamefont {Liang}, \citenamefont {Tseng} \emph
  {et~al.}}]{ozcan:2018:acs}%
  \BibitemOpen
  \bibfield  {author} {\bibinfo {author} {\bibfnamefont {Y.}~\bibnamefont
  {Rivenson}}, \bibinfo {author} {\bibfnamefont {H.}~\bibnamefont
  {Ceylan~Koydemir}}, \bibinfo {author} {\bibfnamefont {H.}~\bibnamefont
  {Wang}}, \bibinfo {author} {\bibfnamefont {Z.}~\bibnamefont {Wei}}, \bibinfo
  {author} {\bibfnamefont {Z.}~\bibnamefont {Ren}}, \bibinfo {author}
  {\bibfnamefont {H.}~\bibnamefont {Gunayd{\i}n}}, \bibinfo {author}
  {\bibfnamefont {Y.}~\bibnamefont {Zhang}}, \bibinfo {author} {\bibfnamefont
  {Z.}~\bibnamefont {Gorocs}}, \bibinfo {author} {\bibfnamefont
  {K.}~\bibnamefont {Liang}}, \bibinfo {author} {\bibfnamefont
  {D.}~\bibnamefont {Tseng}},  \emph {et~al.},\ }\href@noop {} {\bibfield
  {journal} {\bibinfo  {journal} {ACS Photonics}\ }\textbf {\bibinfo {volume}
  {5}},\ \bibinfo {pages} {2354} (\bibinfo {year}
  {2018}{\natexlab{b}})}\BibitemShut {NoStop}%
\bibitem [{\citenamefont {Li}\ \emph {et~al.}(2018{\natexlab{a}})\citenamefont
  {Li}, \citenamefont {Deng}, \citenamefont {Lee}, \citenamefont {Sinha},\ and\
  \citenamefont {Barbastathis}}]{barbastathis:2018:optica}%
  \BibitemOpen
  \bibfield  {author} {\bibinfo {author} {\bibfnamefont {S.}~\bibnamefont
  {Li}}, \bibinfo {author} {\bibfnamefont {M.}~\bibnamefont {Deng}}, \bibinfo
  {author} {\bibfnamefont {J.}~\bibnamefont {Lee}}, \bibinfo {author}
  {\bibfnamefont {A.}~\bibnamefont {Sinha}}, \ and\ \bibinfo {author}
  {\bibfnamefont {G.}~\bibnamefont {Barbastathis}},\ }\href@noop {} {\bibfield
  {journal} {\bibinfo  {journal} {Optica}\ }\textbf {\bibinfo {volume} {5}},\
  \bibinfo {pages} {803} (\bibinfo {year} {2018}{\natexlab{a}})}\BibitemShut
  {NoStop}%
\bibitem [{\citenamefont {Borhani}\ \emph {et~al.}(2018)\citenamefont
  {Borhani}, \citenamefont {Kakkava}, \citenamefont {Moser},\ and\
  \citenamefont {Psaltis}}]{psaltis:2018:optica}%
  \BibitemOpen
  \bibfield  {author} {\bibinfo {author} {\bibfnamefont {N.}~\bibnamefont
  {Borhani}}, \bibinfo {author} {\bibfnamefont {E.}~\bibnamefont {Kakkava}},
  \bibinfo {author} {\bibfnamefont {C.}~\bibnamefont {Moser}}, \ and\ \bibinfo
  {author} {\bibfnamefont {D.}~\bibnamefont {Psaltis}},\ }\href@noop {}
  {\bibfield  {journal} {\bibinfo  {journal} {Optica}\ }\textbf {\bibinfo
  {volume} {5}},\ \bibinfo {pages} {960} (\bibinfo {year} {2018})}\BibitemShut
  {NoStop}%
\bibitem [{\citenamefont {Li}\ \emph {et~al.}(2018{\natexlab{b}})\citenamefont
  {Li}, \citenamefont {Xue},\ and\ \citenamefont {Tian}}]{tian:2018:optica}%
  \BibitemOpen
  \bibfield  {author} {\bibinfo {author} {\bibfnamefont {Y.}~\bibnamefont
  {Li}}, \bibinfo {author} {\bibfnamefont {Y.}~\bibnamefont {Xue}}, \ and\
  \bibinfo {author} {\bibfnamefont {L.}~\bibnamefont {Tian}},\ }\href@noop {}
  {\bibfield  {journal} {\bibinfo  {journal} {Optica}\ }\textbf {\bibinfo
  {volume} {5}},\ \bibinfo {pages} {1181} (\bibinfo {year}
  {2018}{\natexlab{b}})}\BibitemShut {NoStop}%
\bibitem [{\citenamefont {Turpin}\ \emph {et~al.}(2018)\citenamefont {Turpin},
  \citenamefont {Vishniakou},\ and\ \citenamefont {Seelig}}]{turpin:2018:oe}%
  \BibitemOpen
  \bibfield  {author} {\bibinfo {author} {\bibfnamefont {A.}~\bibnamefont
  {Turpin}}, \bibinfo {author} {\bibfnamefont {I.}~\bibnamefont {Vishniakou}},
  \ and\ \bibinfo {author} {\bibfnamefont {J.~D.}\ \bibnamefont {Seelig}},\
  }\href@noop {} {\bibfield  {journal} {\bibinfo  {journal} {Optics Express}\
  }\textbf {\bibinfo {volume} {26}},\ \bibinfo {pages} {30911} (\bibinfo {year}
  {2018})}\BibitemShut {NoStop}%
\bibitem [{\citenamefont {Rahmani}\ \emph {et~al.}(2018)\citenamefont
  {Rahmani}, \citenamefont {Loterie}, \citenamefont {Konstantinou},
  \citenamefont {Psaltis},\ and\ \citenamefont {Moser}}]{moser:2018:lsa}%
  \BibitemOpen
  \bibfield  {author} {\bibinfo {author} {\bibfnamefont {B.}~\bibnamefont
  {Rahmani}}, \bibinfo {author} {\bibfnamefont {D.}~\bibnamefont {Loterie}},
  \bibinfo {author} {\bibfnamefont {G.}~\bibnamefont {Konstantinou}}, \bibinfo
  {author} {\bibfnamefont {D.}~\bibnamefont {Psaltis}}, \ and\ \bibinfo
  {author} {\bibfnamefont {C.}~\bibnamefont {Moser}},\ }\href@noop {}
  {\bibfield  {journal} {\bibinfo  {journal} {Light: Science \& Applications}\
  }\textbf {\bibinfo {volume} {7}},\ \bibinfo {pages} {69} (\bibinfo {year}
  {2018})}\BibitemShut {NoStop}%
\bibitem [{\citenamefont {Caramazza}\ \emph {et~al.}(2019)\citenamefont
  {Caramazza}, \citenamefont {Moran}, \citenamefont {Murray-Smith},\ and\
  \citenamefont {Faccio}}]{caramazza:2019:natcomms}%
  \BibitemOpen
  \bibfield  {author} {\bibinfo {author} {\bibfnamefont {P.}~\bibnamefont
  {Caramazza}}, \bibinfo {author} {\bibfnamefont {O.}~\bibnamefont {Moran}},
  \bibinfo {author} {\bibfnamefont {R.}~\bibnamefont {Murray-Smith}}, \ and\
  \bibinfo {author} {\bibfnamefont {D.}~\bibnamefont {Faccio}},\ }\href@noop {}
  {\bibfield  {journal} {\bibinfo  {journal} {Nature Communications}\ }\textbf
  {\bibinfo {volume} {10}},\ \bibinfo {pages} {2029} (\bibinfo {year}
  {2019})}\BibitemShut {NoStop}%
\bibitem [{\citenamefont {Caramazza}\ \emph {et~al.}(2018)\citenamefont
  {Caramazza}, \citenamefont {Boccolini}, \citenamefont {Buschek},
  \citenamefont {Hullin}, \citenamefont {Higham}, \citenamefont {Henderson},
  \citenamefont {Murray-Smith},\ and\ \citenamefont
  {Faccio}}]{caramazza:2018:scirep}%
  \BibitemOpen
  \bibfield  {author} {\bibinfo {author} {\bibfnamefont {P.}~\bibnamefont
  {Caramazza}}, \bibinfo {author} {\bibfnamefont {A.}~\bibnamefont
  {Boccolini}}, \bibinfo {author} {\bibfnamefont {D.}~\bibnamefont {Buschek}},
  \bibinfo {author} {\bibfnamefont {M.}~\bibnamefont {Hullin}}, \bibinfo
  {author} {\bibfnamefont {C.~F.}\ \bibnamefont {Higham}}, \bibinfo {author}
  {\bibfnamefont {R.}~\bibnamefont {Henderson}}, \bibinfo {author}
  {\bibfnamefont {R.}~\bibnamefont {Murray-Smith}}, \ and\ \bibinfo {author}
  {\bibfnamefont {D.}~\bibnamefont {Faccio}},\ }\href@noop {} {\bibfield
  {journal} {\bibinfo  {journal} {Scientific Reports}\ }\textbf {\bibinfo
  {volume} {8}},\ \bibinfo {pages} {11945} (\bibinfo {year}
  {2018})}\BibitemShut {NoStop}%
\bibitem [{\citenamefont {Faccio}\ \emph {et~al.}(2020)\citenamefont {Faccio},
  \citenamefont {Velten},\ and\ \citenamefont {Wetzstein}}]{NLOSreview}%
  \BibitemOpen
  \bibfield  {author} {\bibinfo {author} {\bibfnamefont {D.}~\bibnamefont
  {Faccio}}, \bibinfo {author} {\bibfnamefont {A.}~\bibnamefont {Velten}}, \
  and\ \bibinfo {author} {\bibfnamefont {G.}~\bibnamefont {Wetzstein}},\
  }\href@noop {} {\bibfield  {journal} {\bibinfo  {journal} {Nat. Rev. Phys.}\
  }\textbf {\bibinfo {volume} {2}},\ \bibinfo {pages} {318--327} (\bibinfo
  {year} {2020})}\BibitemShut {NoStop}%
\bibitem [{\citenamefont {Gariepy}\ \emph
  {et~al.}(2016{\natexlab{b}})\citenamefont {Gariepy}, \citenamefont
  {Tonolini}, \citenamefont {Henderson}, \citenamefont {Leach},\ and\
  \citenamefont {Faccio}}]{gariepy2016detection}%
  \BibitemOpen
  \bibfield  {author} {\bibinfo {author} {\bibfnamefont {G.}~\bibnamefont
  {Gariepy}}, \bibinfo {author} {\bibfnamefont {F.}~\bibnamefont {Tonolini}},
  \bibinfo {author} {\bibfnamefont {R.}~\bibnamefont {Henderson}}, \bibinfo
  {author} {\bibfnamefont {J.}~\bibnamefont {Leach}}, \ and\ \bibinfo {author}
  {\bibfnamefont {D.}~\bibnamefont {Faccio}},\ }\href@noop {} {\bibfield
  {journal} {\bibinfo  {journal} {Nature Photonics}\ }\textbf {\bibinfo
  {volume} {10}},\ \bibinfo {pages} {23} (\bibinfo {year}
  {2016}{\natexlab{b}})}\BibitemShut {NoStop}%
\bibitem [{\citenamefont {O'Toole}\ \emph {et~al.}(2018)\citenamefont
  {O'Toole}, \citenamefont {Lindell},\ and\ \citenamefont
  {Wetzstein}}]{o2018confocal}%
  \BibitemOpen
  \bibfield  {author} {\bibinfo {author} {\bibfnamefont {M.}~\bibnamefont
  {O'Toole}}, \bibinfo {author} {\bibfnamefont {D.~B.}\ \bibnamefont
  {Lindell}}, \ and\ \bibinfo {author} {\bibfnamefont {G.}~\bibnamefont
  {Wetzstein}},\ }\href@noop {} {\bibfield  {journal} {\bibinfo  {journal}
  {Nature}\ }\textbf {\bibinfo {volume} {555}},\ \bibinfo {pages} {338}
  (\bibinfo {year} {2018})}\BibitemShut {NoStop}%
\bibitem [{\citenamefont {Liu}\ \emph {et~al.}(2019)\citenamefont {Liu},
  \citenamefont {Guill{\'e}n}, \citenamefont {La~Manna}, \citenamefont {Nam},
  \citenamefont {Reza}, \citenamefont {Le}, \citenamefont {Jarabo},
  \citenamefont {Gutierrez},\ and\ \citenamefont {Velten}}]{liu2019non}%
  \BibitemOpen
  \bibfield  {author} {\bibinfo {author} {\bibfnamefont {X.}~\bibnamefont
  {Liu}}, \bibinfo {author} {\bibfnamefont {I.}~\bibnamefont {Guill{\'e}n}},
  \bibinfo {author} {\bibfnamefont {M.}~\bibnamefont {La~Manna}}, \bibinfo
  {author} {\bibfnamefont {J.~H.}\ \bibnamefont {Nam}}, \bibinfo {author}
  {\bibfnamefont {S.~A.}\ \bibnamefont {Reza}}, \bibinfo {author}
  {\bibfnamefont {T.~H.}\ \bibnamefont {Le}}, \bibinfo {author} {\bibfnamefont
  {A.}~\bibnamefont {Jarabo}}, \bibinfo {author} {\bibfnamefont
  {D.}~\bibnamefont {Gutierrez}}, \ and\ \bibinfo {author} {\bibfnamefont
  {A.}~\bibnamefont {Velten}},\ }\href@noop {} {\bibfield  {journal} {\bibinfo
  {journal} {Nature}\ }\textbf {\bibinfo {volume} {572}},\ \bibinfo {pages}
  {620} (\bibinfo {year} {2019})}\BibitemShut {NoStop}%
\bibitem [{\citenamefont {Iseringhausen}\ and\ \citenamefont
  {Hullin}(2020)}]{IseringhausenNLOS2020}%
  \BibitemOpen
  \bibfield  {author} {\bibinfo {author} {\bibfnamefont {J.}~\bibnamefont
  {Iseringhausen}}\ and\ \bibinfo {author} {\bibfnamefont {M.~B.}\ \bibnamefont
  {Hullin}},\ }\href {\doibase 10.1145/3368314} {\bibfield  {journal} {\bibinfo
   {journal} {ACM Trans. Graph.}\ }\textbf {\bibinfo {volume} {39}} (\bibinfo
  {year} {2020}),\ 10.1145/3368314}\BibitemShut {NoStop}%
\bibitem [{\citenamefont {Galindo}\ \emph {et~al.}(2019)\citenamefont
  {Galindo}, \citenamefont {Marco}, \citenamefont {O'Toole}, \citenamefont
  {Wetzstein}, \citenamefont {Gutierrez},\ and\ \citenamefont
  {Jarabo}}]{galindo19-NLOSDataset}%
  \BibitemOpen
  \bibfield  {author} {\bibinfo {author} {\bibfnamefont {M.}~\bibnamefont
  {Galindo}}, \bibinfo {author} {\bibfnamefont {J.}~\bibnamefont {Marco}},
  \bibinfo {author} {\bibfnamefont {M.}~\bibnamefont {O'Toole}}, \bibinfo
  {author} {\bibfnamefont {G.}~\bibnamefont {Wetzstein}}, \bibinfo {author}
  {\bibfnamefont {D.}~\bibnamefont {Gutierrez}}, \ and\ \bibinfo {author}
  {\bibfnamefont {A.}~\bibnamefont {Jarabo}},\ }\href
  {https://graphics.unizar.es/nlos} {\enquote {\bibinfo {title} {A dataset for
  benchmarking time-resolved non-line-of-sight imaging},}\ } (\bibinfo {year}
  {2019})\BibitemShut {NoStop}%
\bibitem [{\citenamefont {Klein}\ \emph {et~al.}(2019)\citenamefont {Klein},
  \citenamefont {Laurenzis}, \citenamefont {Michels},\ and\ \citenamefont
  {Hullin}}]{hullin-dataset}%
  \BibitemOpen
  \bibfield  {author} {\bibinfo {author} {\bibfnamefont {J.}~\bibnamefont
  {Klein}}, \bibinfo {author} {\bibfnamefont {M.}~\bibnamefont {Laurenzis}},
  \bibinfo {author} {\bibfnamefont {D.~L.}\ \bibnamefont {Michels}}, \ and\
  \bibinfo {author} {\bibfnamefont {M.~B.}\ \bibnamefont {Hullin}},\ }\href
  {https://nlos.cs.uni-bonn.de/} {\enquote {\bibinfo {title} {Nlos
  benchmark},}\ } (\bibinfo {year} {2019})\BibitemShut {NoStop}%
\bibitem [{\citenamefont {Goodfellow}\ \emph {et~al.}(2016)\citenamefont
  {Goodfellow}, \citenamefont {Bengio},\ and\ \citenamefont
  {Courville}}]{goodfellow:2016:book}%
  \BibitemOpen
  \bibfield  {author} {\bibinfo {author} {\bibfnamefont {I.}~\bibnamefont
  {Goodfellow}}, \bibinfo {author} {\bibfnamefont {Y.}~\bibnamefont {Bengio}},
  \ and\ \bibinfo {author} {\bibfnamefont {A.}~\bibnamefont {Courville}},\
  }\href@noop {} {\emph {\bibinfo {title} {Deep learning}}}\ (\bibinfo
  {publisher} {MIT press},\ \bibinfo {year} {2016})\BibitemShut {NoStop}%
\bibitem [{\citenamefont {Chollet}(2015)}]{keras}%
  \BibitemOpen
  \bibfield  {author} {\bibinfo {author} {\bibfnamefont {F.}~\bibnamefont
  {Chollet}},\ }\href@noop {} {\enquote {\bibinfo {title} {keras},}\ }\bibinfo
  {howpublished} {\url{https://github.com/fchollet/keras}} (\bibinfo {year}
  {2015})\BibitemShut {NoStop}%
\bibitem [{\citenamefont {Abadi~\textit{et al}}(2015)}]{tensorflow2015}%
  \BibitemOpen
  \bibfield  {author} {\bibinfo {author} {\bibfnamefont {M.}~\bibnamefont
  {Abadi~\textit{et al}}},\ }\href {http://tensorflow.org/} {\enquote {\bibinfo
  {title} {{TensorFlow}: Large-scale machine learning on heterogeneous
  systems},}\ } (\bibinfo {year} {2015})\BibitemShut {NoStop}%
\bibitem [{\citenamefont {Sanzaro}\ \emph {et~al.}(2018)\citenamefont
  {Sanzaro}, \citenamefont {Gattari}, \citenamefont {Villa}, \citenamefont
  {Croce},\ and\ \citenamefont {Zappa}}]{sanzaro:2018:IEEE}%
  \BibitemOpen
  \bibfield  {author} {\bibinfo {author} {\bibfnamefont {M.}~\bibnamefont
  {Sanzaro}}, \bibinfo {author} {\bibfnamefont {P.}~\bibnamefont {Gattari}},
  \bibinfo {author} {\bibfnamefont {F.}~\bibnamefont {Villa}}, \bibinfo
  {author} {\bibfnamefont {G.}~\bibnamefont {Croce}}, \ and\ \bibinfo {author}
  {\bibfnamefont {F.}~\bibnamefont {Zappa}},\ }\href@noop {} {\bibfield
  {journal} {\bibinfo  {journal} {IEEE Journal of Selected Topics in Quantum
  Electronics}\ }\textbf {\bibinfo {volume} {24}},\ \bibinfo {pages} {1}
  (\bibinfo {year} {2018})}\BibitemShut {NoStop}%
\bibitem [{\citenamefont {Wang}\ \emph {et~al.}(2004)\citenamefont {Wang},
  \citenamefont {Bovik}, \citenamefont {Sheikh},\ and\ \citenamefont
  {Simoncelli}}]{simoncelli:2004:IEEE}%
  \BibitemOpen
  \bibfield  {author} {\bibinfo {author} {\bibfnamefont {Z.}~\bibnamefont
  {Wang}}, \bibinfo {author} {\bibfnamefont {A.~C.}\ \bibnamefont {Bovik}},
  \bibinfo {author} {\bibfnamefont {H.~R.}\ \bibnamefont {Sheikh}}, \ and\
  \bibinfo {author} {\bibfnamefont {E.~P.}\ \bibnamefont {Simoncelli}},\
  }\href@noop {} {\bibfield  {journal} {\bibinfo  {journal} {IEEE transactions
  on image processing}\ }\textbf {\bibinfo {volume} {13}},\ \bibinfo {pages}
  {600} (\bibinfo {year} {2004})}\BibitemShut {NoStop}%
\end{thebibliography}
\end{document}